\documentclass[conference]{IEEEtran}
\IEEEoverridecommandlockouts

\usepackage{enumitem}
\usepackage{array}
\usepackage{wrapfig}
\usepackage{amsmath,amsfonts}
\usepackage[noend]{algpseudocode}
\usepackage{graphicx}
\usepackage{textcomp}
\usepackage{float}
\usepackage{listings}
\usepackage{xspace}
\usepackage{multirow}
\usepackage{amsthm}

\usepackage{balance}
\usepackage{algorithm}
\usepackage{algpseudocode}
\usepackage{colortbl}

\usepackage[skins]{tcolorbox}
\usepackage{xcolor,pifont}
\usepackage{multicol}

\usepackage{listings}
\usepackage{xcolor}
\usepackage{array}
\usepackage{tcolorbox}
\usepackage[normalem]{ulem}
\usepackage{wrapfig}
\usepackage{placeins}
\usepackage{tabularx}
\usepackage{makecell}
\newcolumntype{C}{>{\centering\arraybackslash}X} 

\usepackage{enumitem}

\usepackage{caption}

\usepackage{enumitem}
\usepackage{url}
\usepackage{subcaption}

\usepackage{booktabs}
\usepackage{multirow}

\newcommand*\colourcheck[1]{
	\expandafter\newcommand\csname #1check\endcsname{\textcolor{#1}{\ding{52}}}
}
\colourcheck{blue}
\colourcheck{green}
\colourcheck{red}
\newtcolorbox{boxB}[2][]{
  enhanced,colback=white,colframe=black,coltitle=black,
  sharp corners,
  toprule=1.0pt,
  rightrule=0.3pt,
  leftrule=0pt,
  bottomrule=0pt,
  fonttitle=\itshape\scshape\large,
  left=0pt,right=5pt,top=5pt,bottom=3pt,
  attach boxed title to top right={yshift=-0.3\baselineskip-0.4pt,xshift=-5mm},
  boxed title style={tile,size=minimal,left=0.2mm,right=0.5mm,
    colback=white,before upper=\strut},
  title=#2,#1
}

\newcommand{\pair}[2]{
#1/\ifnum#2>#1\colorbox{green!20}{#2}\else #2\fi}

\definecolor{DarkGreen}{rgb}{0, 0.5, 0}
\definecolor{DarkRed}{rgb}{0.8, 0, 0}
\definecolor{siggreen}{RGB}{198,224,180}
\definecolor{nonsigred}{RGB}{230,184,183}

\newcommand{\high}[1]{\textcolor{DarkGreen}{#1}}
\newcommand{\low}[1]{\textcolor{DarkRed}{#1}}

\newcolumntype{C}{>{\centering\arraybackslash}X}

\AtBeginDocument{
  \providecommand\BibTeX{{
      Bib\TeX}}}

\lstdefinelanguage{Python}{
    morekeywords={class, def, return, try, except, raise, from},
    keywordstyle=\color{blue},
    stringstyle=\color{green},
    commentstyle=\color{gray},
    morecomment=[l]{\#},
}

\lstdefinelanguage{diff}{
    morecomment=[f][\color{green}]{+},
    morecomment=[f][\color{red}]{-},
    morecomment=[f][\color{blue}]{@@},
}

\lstdefinelanguage{errorlog}{
    morecomment=[f][\color{red}]{E},
    morecomment=[f][\color{magenta}]{?},
}

\newcolumntype{G}{>{\columncolor{green!12}}c}

\usepackage{listings}
\usepackage{xspace}

\usepackage[T1]{fontenc}

\usepackage[utf8]{inputenc}

\usepackage{microtype}

\usepackage{graphicx}

\usepackage{subcaption}

\usepackage{soul}
\usepackage{framed}
\usepackage{multirow}
\usepackage{siunitx}

\definecolor{custom-blue}{rgb}{0,0,0}

\newboolean{showcomments}
\setboolean{showcomments}{true}

\ifthenelse{\boolean{showcomments}}
 { \newcommand{\mynote}[2]{
      \fbox{\bfseries\sffamily\scriptsize#1}
        {\small$\blacktriangleright$\textsf{\emph{#2}}$\blacktriangleleft$}}}
        { \newcommand{\mynote}[2]{}}

\newcolumntype{L}[1]{>{\raggedright\arraybackslash}p{#1}}

\newcommand{\code}[1]{{\small\texttt{#1}}}
\usepackage{amsthm}
\definecolor{dkgreen}{rgb}{0,0.6,0}
\definecolor{gray}{rgb}{0.5,0.5,0.5}
\definecolor{lightgray}{rgb}{211, 211, 211}
\definecolor{mauve}{rgb}{0.58,0,0.82}

\definecolor{lightred}{rgb}{1,0.85,0.85}
\definecolor{lightblue}{rgb}{0.85,0.92,1}
\definecolor{lightgreen}{rgb}{0.85,1,0.85}

\definecolor{custom-red}{rgb}{1,0,0}

\definecolor{custom-blue}{rgb}{0,0,1}

\definecolor{c1}{HTML}{f4cccc}
\definecolor{c2}{HTML}{f5cdcd}
\definecolor{c3}{HTML}{fffcfc}
\definecolor{c4}{HTML}{ffffff}
\definecolor{c5}{HTML}{ffffff}
\definecolor{c6}{HTML}{fffdfd}
\definecolor{c7}{HTML}{f5cfcf}
\definecolor{c8}{HTML}{fffbfb}
\definecolor{c9}{HTML}{ffffff}
\definecolor{c10}{HTML}{fffdfd}
\definecolor{c11}{HTML}{fefafa}
\definecolor{c12}{HTML}{fef7f7}
\definecolor{c13}{HTML}{ffffff}
\definecolor{c14}{HTML}{fffefe}
\definecolor{c15}{HTML}{ffffff}
\definecolor{c16}{HTML}{fefafa}
\definecolor{c17}{HTML}{fdf3f3}
\definecolor{c18}{HTML}{fffefe}
\definecolor{c19}{HTML}{fdf5f5}
\definecolor{c20}{HTML}{ffffff}

\definecolor{acc6}{HTML}{C62027}
\definecolor{acc7}{HTML}{CA2427}
\definecolor{acc9}{HTML}{D12E27}

\definecolor{acc35}{HTML}{F8B55E}
\definecolor{acc36}{HTML}{F9B95C}
\definecolor{acc37}{HTML}{F9BD5D}
\definecolor{acc38}{HTML}{FAC15E}
\definecolor{acc39}{HTML}{FBC55F}
\definecolor{acc40}{HTML}{FCC960}
\definecolor{acc41}{HTML}{FDCD62}
\definecolor{acc43}{HTML}{FDD76B}
\definecolor{acc45}{HTML}{FEE074}
\definecolor{acc46}{HTML}{FEE47A}
\definecolor{acc48}{HTML}{F5F9A9}
\definecolor{acc50}{HTML}{E9F6A1}

\definecolor{acc54}{HTML}{CDEB86}
\definecolor{acc56}{HTML}{BFE47A}
\definecolor{acc57}{HTML}{B6E075}
\definecolor{acc64}{HTML}{84CA66}
\definecolor{acc3}{HTML}{C62027}
\definecolor{acc4}{HTML}{C62027}
\definecolor{acc5}{HTML}{C62027}
\definecolor{acc10}{HTML}{D53227}
\definecolor{acc11}{HTML}{D93627}
\definecolor{acc12}{HTML}{DD3A27}
\definecolor{acc13}{HTML}{E13E27}
\definecolor{acc16}{HTML}{E94A27}
\definecolor{acc17}{HTML}{ED4E27}
\definecolor{acc20}{HTML}{F15A27}
\definecolor{acc23}{HTML}{F56627}
\definecolor{acc33}{HTML}{F7AF5E}
\definecolor{acc34}{HTML}{F8B15E}
\definecolor{acc44}{HTML}{FEE074}
\definecolor{acc52}{HTML}{DAF192}
\newcommand{\acc}[1]{\cellcolor{acc#1}#1\%}

\newlength{\cellsize}

\def\BibTeX{{\rm B\kern-.05em{\sc i\kern-.025em b}\kern-.08em
    T\kern-.1667em\lower.7ex\hbox{E}\kern-.125emX}}

\usepackage{tikz}

\makeatletter
\newcommand{\linebreakand}{
  \end{@IEEEauthorhalign}
  \hfill\mbox{}\par
  \mbox{}\hfill\begin{@IEEEauthorhalign}
}
\makeatother

\usepackage{lineno}

\begin{document}

\title{Do Machines Struggle Where Humans Do? LLM and Human Comprehension of Obfuscated Code}

\author{\IEEEauthorblockN{Jack Le\textsuperscript{1,*}, Anh H.N. Nguyen\textsuperscript{1,*}, and Tien N. Nguyen\textsuperscript{1}}
\IEEEauthorblockA{\textit{Department of Computer Science} \\
\textit{University of Texas at Dallas}\\
Richardson, Texas, USA \\
\textsuperscript{1}\{jvl210002, Tien.N.Nguyen\}@utdallas.edu}
\thanks{*Jack Van Le and Anh H.N. Nguyen contributed equally to this work.}
}
\maketitle

\begin{abstract}
While code obfuscation impairs human code comprehension, it remains unclear if large language models share these failure modes. Building directly on a recent human study of program comprehension under code obfuscation~\cite{nguyen2026effectcodeobfuscationhuman}, we evaluate whether large language models share the failure modes that obfuscation induces in human programmers. Evaluating several LLMs with five obfuscation tiers using the Block Model, we localize comprehension failures at the atom, block, relational, and macro levels. We find that reasoning-tuned models demonstrate significant alignment with human difficulty patterns across experience levels, whereas instruction and coder-tuned models show near-zero correlation. Chain-of-Thought trace length tracks task difficulty across tasks. Results indicate that performance under control-flow flattening degrades in proportion to state-space complexity, while adversarial identifier renaming disrupts comprehension through the interaction of semantic displacement and identifier-level interference. These findings suggest that reasoning-tuned LLMs approximate human sensitivity to code complexity more effectively than instruction-tuned variants.
\end{abstract}

\section{Introduction}

Code obfuscation preserves program semantics while deliberately making code harder to understand. It is widely used to protect software from reverse engineering, tampering, and unauthorized reuse, but it also provides a useful lens for studying program comprehension itself. Because obfuscation removes or distorts familiar lexical and structural cues without changing program behavior, it creates a controlled setting for testing whether an agent truly understands program logic or instead relies on patterns such as meaningful identifiers, common idioms, or recognizable control-flow structures.

Nguyen et al.~\cite{nguyen2026effectcodeobfuscationhuman} showed that obfuscation disrupts the cues programmers use to form high-level semantic hypotheses, increases cognitive effort, and forces more deliberate reasoning. This raises a fundamental question for code LLMs: do machines understand obfuscated code in ways that align with human understanding, or do they fail differently? If they align, obfuscated code becomes a principled benchmark for human-like reasoning, and robustness would signal genuine comprehension. If they diverge, high benchmark performance may mask a brittle competence that succeeds on standard tasks yet breaks under transformations humans can still reason through.

Motivated by this, we study the alignment between human and machine understanding of obfuscated code, asking whether different LLM classes mirror human difficulty patterns, whether reasoning-tuned models are more aligned than coder or instruction-tuned models, and whether models reproduce the specific failure modes obfuscation induces in humans: effort that scales with task difficulty, and confident misinterpretation under misleading identifiers. Together, these questions move beyond aggregate accuracy to examine the nature of model understanding under controlled distortion.

To answer them, we conduct a mixed empirical study~combining model evaluation and comparison with prior human results~\cite{nguyen2026effectcodeobfuscationhuman}. We evaluate several LLMs on obfuscated versions of two datasets in Python and JavaScript across five obfuscation tiers, including identifier renaming, adversarial renaming, control-flow flattening, and their combination. We analyze both output prediction performance and reasoning traces, and we interpret failures through Schulte's Block Model of program comprehension~\cite{schulte2008block}, which separates atom-level, block-level, relational, and macro-level understanding. This enables direct comparison between model behavior and previously reported human difficulty patterns under the same obfuscation types~\cite{nguyen2026effectcodeobfuscationhuman}. 

% Our study builds directly on the experimental paradigm of Nguyen \textit{et al.}~\cite{nguyen2026effectcodeobfuscationhuman}: we adopt their obfuscation tiers, their output-prediction protocol, and their human comprehension data, and extend this paradigm to large language models. The human study contributes the difficulty patterns against which we measure; the novel contribution of this work is the model-side evaluation and the human--machine alignment analysis built on top of it.

This paper makes the following~contributions:

\begin{itemize} [leftmargin=0.7em, labelsep=0.2em]
    \item We formulate the problem of measuring human--machine alignment in obfuscated-code understanding, following Nguyen \textit{et al.}~\cite{nguyen2026effectcodeobfuscationhuman} as the human baseline for our evaluation.
    \item We conduct a systematic study of LLMs across multiple obfuscation tiers in Python and JavaScript, comparing their behavior against the reported human difficulty patterns.
    \item We analyze model failures using the Block Model and reasoning traces, showing where alignment emerges and where machine understanding diverges from human understanding.
    % \item We provide evidence that reasoning-tuned LLMs are more human-aligned than other model classes, while also showing that this alignment remains incomplete under stronger obfuscation
    \item We report the findings and implications for future studies in both software engineering and software security.
    
    %provide evidence that reasoning-tuned LLMs exhibit significantly greater cognitive alignment with human difficulty patterns than coder or instruction-tuned models — whose failure modes appear driven by processes fundamentally distinct from human reasoning — and that this alignment is experience-stratified, correlating most strongly with intermediate and expert participants.

\end{itemize}

%We evaluate twelve models across two experiments and five obfuscation tiers using Carsten Schulte's Block Model to localize comprehension failures at the atom, block, relational, and macro levels. Drawing on a parallel human study, we find that reasoning-tuned models align significantly with human difficulty patterns, while coder and instruction-tuned models show near-zero correlation. 
\section{Background}

\subsection{Code Obfuscation}
Prior work categorizes obfuscation techniques by transformation properties, most notably distinguishing layout-based from control-flow-based obfuscations~\cite{collberg1997taxonomy}. Identifier renaming is a layout-based obfuscation, while control-flow flattening is a control-flow-based obfuscation~\cite{ceccato2008towards}. We adopt these obfuscation levels to align with Nguyen {\em et al.}~\cite{nguyen2026effectcodeobfuscationhuman}'s human study on code understandability under obfuscation. 

\textbf{L1: Identifier Renaming}. Identifier renaming is a layout-based obfuscation that replaces meaningful identifiers (e.g., function/variable names) with short, incoherent, or minimally informative names. This disrupts the semantic cues normally provided by identifiers, making it harder to~understand. 

\textbf{L1b: Adversarial Renaming}. We use L1b, a variant of L1 to capture a distinct effect of identifier obfuscation.~While L1 removes semantic cues, L1b replaces identifiers with semantically meaningful but misleading names in the new~context. This semantic mismatch can induce incorrect mental models of the code, leading to high-confidence but erroneous~interpretations.

\textbf{L2: Control-Flow Alteration}. This obfuscates the execution by decoupling control logic from its original syntactic structure. It decomposes code into smaller blocks and introduces artificial control constructs (e.g., dispatchers or indirect jumps), making the original execution order difficult to reconstruct~\cite{laszlo2007cff}.

\textbf{L3: Combination of L1 and L2}. L3 combines identifier renaming and control-flow alteration. As the most complex setting, L3 serves as an upper bound on obfuscation difficulty for the assessment of both human and model code understanding.

\subsection{Human Understanding of Obfuscated Code}
\label{sec:human-study}

Nguyen \emph{et al.}~\cite{nguyen2026effectcodeobfuscationhuman} studied human program comprehension under code obfuscation. They recruited 50 undergraduate computer science students to perform output-prediction tasks under controlled conditions. Each participant was asked to inspect a Python or JavaScript function together with a concrete input and predict the exact output, while the study recorded correctness, response time, and self-reported programming experience. The dataset contains 20 function-level output-prediction tasks from HumanEval-X, with 10 JavaScript snippets and 10 Python snippets with cyclomatic complexity between 4 and 8 and fewer than 15 lines, making them non-trivial but still feasible for manual reasoning. Each snippet is instantiated across~five obfuscation tiers L0-L3. We use this dataset to enable direct comparison between LLM behavior and human patterns. Our work extends this study from human participants to large language models, reusing its tasks, obfuscation tiers, and reported difficulty patterns.

\section{Research Questions}

\begin{figure}
    \centering
    \includegraphics[width=\linewidth]{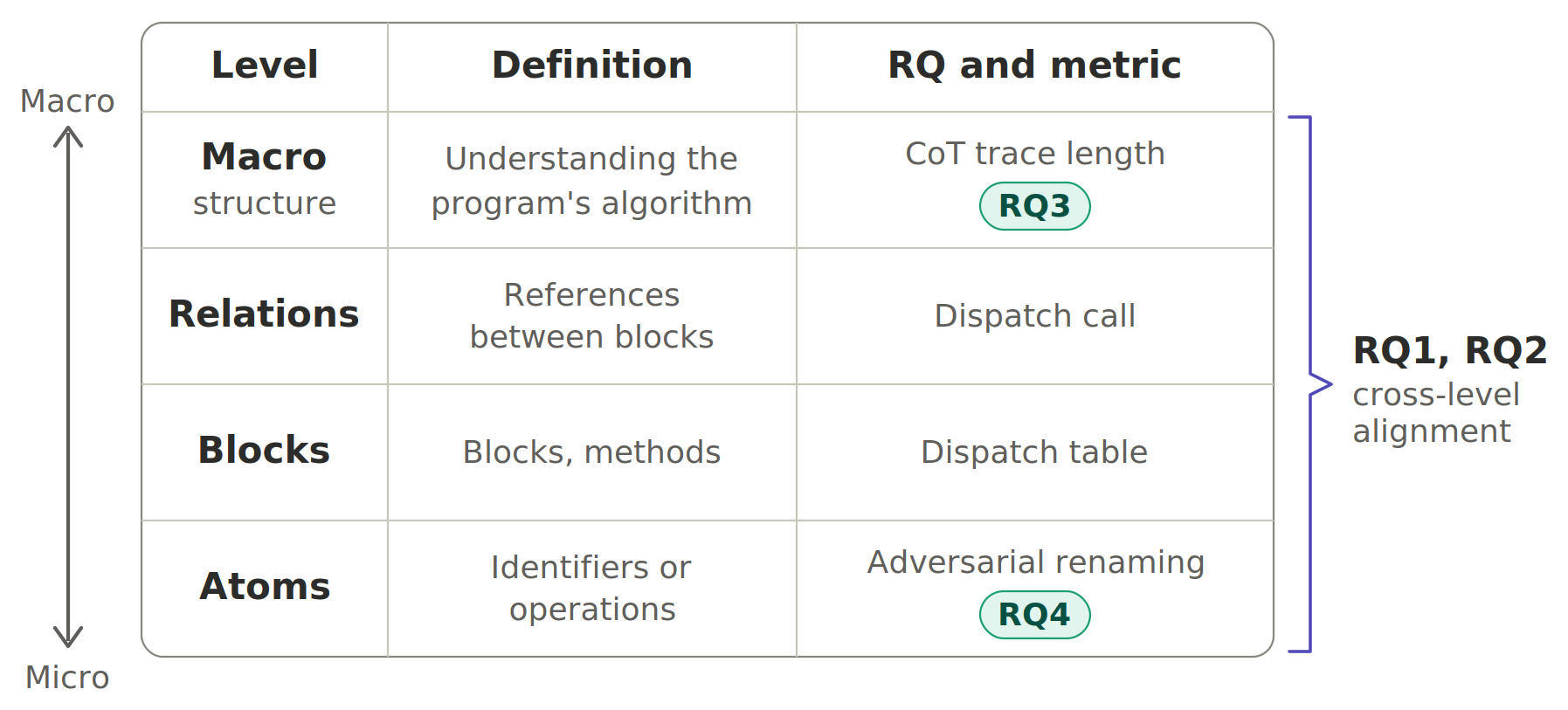}
    \vspace{-21pt}
    \caption{Block Model Schema}
    \label{fig:block}
    \vspace{-5pt}
\end{figure}

We organize our research questions using Carsten Schulte's Block 
Model~\cite{schulte2008block} (Fig.~\ref{fig:block}), which 
organizes program comprehension into four hierarchical levels: 
\textit{Atoms} (individual language elements), \textit{Blocks} 
(semantically meaningful regions), \textit{Relations} 
(dependencies and data flow between blocks), and \textit{Macro 
Structure} (the overall algorithm and purpose). We use output 
prediction as a proxy for understanding as in Nguyen {\em et al.}~\cite{nguyen2026effectcodeobfuscationhuman}. The Block Model allows us to localize where 
comprehension breaks down and connect each failure mode to a 
specific structural property of obfuscated code.

Together these questions probe two levels of alignment. RQ1–RQ2 ask whether models find the same
tasks hard as humans (outcome-level), spanning all Block-Model levels — including the relational
level, where control-flow obfuscation forces models to resolve dispatcher-mediated inter-block
dependencies. RQ3–RQ4 ask whether models reproduce the failure modes obfuscation induces in
humans (mechanism-level): effort scaling at the macro level, proxied by CoT trace length (RQ3),
and confident misinterpretation under misleading identifiers at the atom level (RQ4).

\noindent\textbf{RQ1: To what extent do LLMs succeed and fail on 
the same obfuscated code tasks as human participants?}
Nguyen {\em et al.}~\cite{nguyen2026effectcodeobfuscationhuman} shows that 
obfuscation shifts human reasoning from System 1 to System 2. We 
investigate whether LLMs exhibit the same task-level difficulty 
patterns as humans. 

\noindent \textbf{RQ2: Do different model architectures and sizes align with human expertise levels?} While RQ1 examines whether models track human difficulty patterns, RQ2 investigates if any model's reasoning profile resembles that of a particular experience bracket. Humans with varied experience employ qualitatively different strategies under obfuscation. We investigate whether model size or architecture produces analogous stratification.

\noindent \textbf{RQ3: Can CoT Reasoning length traces measure an LLM's reasoning and task difficulty?}  

At the macro level of the Block Model, we use CoT trace length as a proxy for mental effort. The length of a model's CoT tokens has been shown to correlate with the time humans take to solve the same problems~\cite{devarda2025cost}. Building on this, we investigate whether CoT trace length serves as a reliable indicator of reasoning depth and inherent task difficulty, or whether longer traces reflect model confusion rather than productive deduction. Because the human study records response time as a measure of effort~\cite{nguyen2026effectcodeobfuscationhuman}, this lets us compare directly how human and model effort scale with obfuscation.

\noindent \textbf{RQ4: Does adversarial renaming cause LLMs to generate high-confidence, incorrect answers due to semantic mis-framing?} Human understanding treats identifiers as semantic beacons, so misleading names can induce confident but~incor\-rect mental models--the failure mode adversarial renaming (L1b) was designed to elicit~\cite{nguyen2026effectcodeobfuscationhuman}. We ask whether models inherit the same atom-level vulnerability. At the atom level, standard and adversarial renaming present different forms of semantic transformation, which may affect models beyond accuracy. We study how adversarial renaming and semantic displacement~relate to accuracy, confidence, and high-confidence incorrect~predictions.

\section{Experimental Methodology}

\subsection{Datasets and Model Selection}

\subsubsection{Datasets} We used two datasets $A$  and $B$: {\bf \em Dataset A} is drawn from Nguyen {\em et al.}'s study~\cite{nguyen2026effectcodeobfuscationhuman} to enable the study on the alignment of human and model understanding as they contain the human labels for understandability (Section~\ref{sec:human-study}). To further study models' understanding, we collected our own {\bf \em dataset B}, comprising 250 snippets in total, with 10 unique questions sampled from each benchmark family (HumanEval-X Python/JavaScript \cite{humanevalx}, CruxEval-X Python/JavaScript \cite{cruxevalx}, LeetCode \cite{leetcodedataset}), instantiated across all five obfuscation tiers (L0–L3).
To mitigate training data contamination, we consider a temporal cutoff based on the LeetCode dataset construction date (05/2025). For problems with multiple input-output pairs, we randomly select a single pair with a fixed seed and reuse it consistently across all tiers. We retain only self-contained, deterministic tasks, excluding cases involving external state, or randomized behavior. Obfuscation tiers L1–L3 are generated using ObfuXtreme~\cite{obfuXtreme25} (Python) and javascript-obfuscator~\cite{javascript_obfuscator_411} (JavaScript), configured to isolate identifier renaming (L1) and control-flow flattening (L2), with L3 combining both.~All obfuscated programs are validated by execution to ensure~functional equivalence with the original code. 

Dataset A is used for RQ1--RQ2 (human--model alignment); 
Dataset B for RQ3--RQ4 (model-based output prediction). 

\subsubsection{Models} Our experimental framework spans diverse architectures and scales, including the Llama, Qwen, DeepSeek, Phi, and SmolLM families. To capture emerging reasoning capabilities, we include models such as SmolLM3-3B and Qwen3-0.6B, which provide explicit “think” modes enabling analysis of System 1 vs. System 2 behavior; both are evaluated in reasoning mode. We also include a small number of post-cutoff models (e.g., SmolLM3-3B, Phi-4-mini-flash-reasoning, and Llama-3.1-8B-Instruct) to improve coverage across model types and architectures. As release dates are an imperfect proxy for training data cutoff, results are interpreted with appropriate caution rather than strict temporal separation.

\begin{table}[t]
\caption{Details of selected models.}
\centering
\footnotesize
\setlength{\tabcolsep}{4pt}
\renewcommand{\arraystretch}{1.15}
\vspace{-6pt}
\begin{tabularx}{\columnwidth}{@{} l >{\raggedright\arraybackslash}X l l @{}}
\toprule
\textbf{Size} & \textbf{Model} & \textbf{Type} & \textbf{Release} \\
\midrule
\textbf{Small}
    & DeepSeek-R1-Distill-Qwen-1.5B \cite{paper-model-deepseek-r1-distill-qwen-1.5b-7b} & Reason & Jan '25 \\
    & Phi-4-mini-flash-reasoning (4B) \cite{paper-model-phi-4-mini-flash-reasoning} & Reason & Jul '25 \\
    & SmolLM3-3B \cite{paper-model-smollm3-3b} & Reason & Jul '25 \\
    & Qwen3-0.6B \cite{paper-model-qwen3-0.6b} & Reason & Apr '25 \\
    & Qwen2.5-0.5B-Instruct \cite{paper-model-qwen2.5-0.5b-instruct} & Instr. & Sep '24 \\
    & Phi-3.5-mini-instruct \cite{paper-model-phi-3.5-mini-instruct} & Instr. & Aug '24 \\
\midrule
\textbf{Med.}
    & DeepSeek-R1-Distill-Qwen-7B \cite{paper-model-deepseek-r1-distill-qwen-1.5b-7b} & Reason & Jan '25 \\
    & Qwen-7B \cite{paper-model-qwen-7b} & Instr. & Sep '23 \\
    & Llama-3.1-8B-Instruct \cite{paper-model-llama3.1-8b-instruct} & Instr. & Jul '25 \\
    & CodeLlama-7B-Instruct-hf \cite{paper-model-codellama-7b-instruct} & Coder & Jul '23 \\
    & Qwen2.5-Coder-7B-Instruct \cite{paper-model-qwen2.5-coder-7b-instruct} & Coder & Sep '24 \\
    & DeepSeek-Coder-6.7B-Instruct \cite{paper-model-deepseekcoder-6.7b} & Coder & Nov '23 \\
\bottomrule
\end{tabularx}
\label{tab:combined_models_no_notes}
\end{table}

\subsection{Prompt Design}
\label{subsec:prompts}

\begin{figure}
    \centering
    \includegraphics[width=1\linewidth]{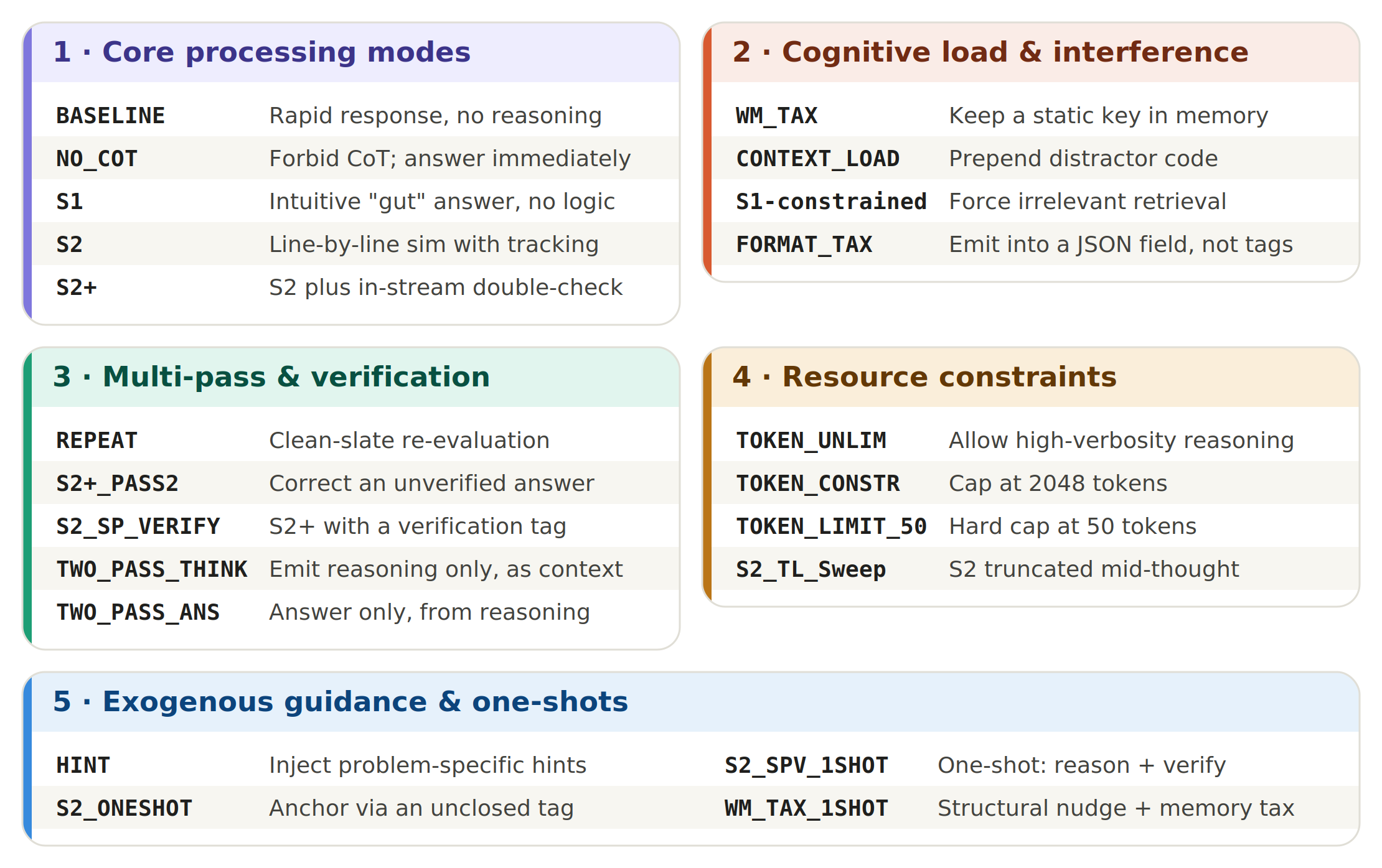}
    \caption{Prompt Design: Different Variations}
    \label{fig:prompt-variations}
\end{figure}

Our prompt suite is designed around five orthogonal axes---reasoning depth, cognitive interference, verification, token budget, and external scaffolding---so that any performance shift can be attributed to a specific manipulation rather than incidental wording. We began from two anchor prompts, a bare output-prediction request (\code{BASELINE}) and an explicit line-by-line simulation request (\code{S2}), and derived the remaining conditions by applying one controlled modification at a time (e.g., a memory tax, a token cap, a one-shot exemplar) while holding the task instruction and answer-extraction format fixed. This yields the prompt conditions summarized in Figure~\ref{fig:prompt-variations}, which probe several dimensions of model reasoning: we vary the \textit{depth of reasoning elicited} (Category~1), from rapid baseline responses to line-by-line simulation (\code{S2}); introduce \textit{cognitive interference} (Category~2), such as holding a static key in memory (\code{WM\_TAX}) or prepending distractor code (\code{CONTEXT\_LOAD}); require \textit{multi-pass verification} (Category~3), where the model re-checks its own output (\code{TWO\_PASS\_THINK}); constrain the \textit{token budget} (Category~4), from unconstrained to a 50-token hard limit, separating reasoning verbosity from answer quality; and supply \textit{exogenous guidance} (Category~5), such as hints or one-shot examples, to probe how external scaffolding interacts with intrinsic reasoning. To keep conditions comparable across models, instruction bodies are identical for all models, with only the official per-model chat template applied and no per-model prompt tuning; reasoning-mode models are run in their native ``think'' mode. Complete prompt text is available on replication package~\cite{anon_data_repo_2026}.

All experiments were done on four 
NVIDIA RTX A6000 GPUs. Dataset A comprises 21 prompt 
conditions (25 total passes, as one condition repeats 5 times 
for self-consistency) $\times$ 20 snippets $\times$ 5 
obfuscation tiers $\times$ 7 models, totaling 17,500 
evaluation runs. Dataset B comprises 5 prompt conditions 
$\times$ 50 snippets $\times$ 5 obfuscation tiers $\times$ 9 
models, totaling 11,250~runs.

\subsection{Evaluation Metrics and Block-Model Measures}
\label{subsec:measures}

We assess comprehension with one primary outcome -- output accuracy --
and a set of measures that localize failures to specific levels of the
Block Model (Fig.~\ref{fig:block}). 

\textbf{Output accuracy (All).} All models perform output prediction
and a response is correct only if its predicted output is an exact match to the ground-truth output obtained by executing the reference program on that input. All comparisons use mixed-effects logistic regression (binomial GLMMs
with crossed random effects for snippet and model); proportions are reported with
Wilson 95\% confidence intervals, $p$-values are Benjamini--Hochberg
FDR-corrected within each research question, and we report effect sizes
(odds ratios, Cliff's $\delta$).

\textbf{Macro level --- CoT trace length (RQ3).} The number of reasoning
tokens generated before the final answer, used as a proxy for mental effort.

\textbf{Relational/block level --- dispatcher complexity (RQ1).} Under
control-flow flattening, original control flow is replaced by
dispatcher constructs. We quantify their complexity with the Python
while-if state count, JavaScript dispatch-call references, and
object-dispatch helpers; these are used in Section~\ref{acc_under_obf}.

\textbf{Atom level --- adversarial-renaming measures (RQ4).} These
measures isolate the
lexical-semantic effects of adversarial renaming. We measure \emph{semantic
distance} at the identifier level using SFR embeddings~\cite{SFREmbeddingModel},
computing a weighted score combining cosine distance with a semantic-shift
term (capturing how much the renamed identifier becomes more semantically
suggestive than the original). Snippet-level distance is the mean over all
aligned identifier pairs, with identifiers embedded in isolation after
normalization to isolate lexical-semantic effects from broader context.
\emph{Adversarial mis-framing} is the additional semantic distance
introduced by $L1b$ relative to $L1$.
\[
\Delta_{\text{misframe}} = D_{L0 \rightarrow L1b} - D_{L0 \rightarrow L1}.
\]
We introduce two measures of confident, localized failure. \emph{High-Confidence
Incorrect} (HCI) flags wrong answers produced at group-normalized confidence
$z_i \ge 1.0$, where $z_i$ is the confidence defined earlier. \emph{Identifier-level disruption} is
captured by the \emph{Identifier Spike Fraction} (ISF), the share of
generation uncertainty concentrated on identifier tokens, binned by $Q_3$
into a \emph{spike regime} $\{\text{low}, \text{med}, \text{high}\}$. A
\emph{spike} is a token position where the model is unusually surprised by
its own generated token relative to the same forward pass; token $t$ is
flagged if
\begin{equation*}
\mathrm{ppl}_t = \exp(-\log p(x_t \mid x_{<t})) >
\mathrm{median}(\mathrm{ppl}) + 3\cdot\mathrm{MAD}(\mathrm{ppl})
\end{equation*}
or falls in the top $2\%$ of per-token perplexities in the trace, where
$\mathrm{MAD} = \mathrm{median}(|\mathrm{ppl}_t - \mathrm{median}(\mathrm{ppl})|)$.

Prior work showed that model probability-based uncertainty signals can serve as effective quality/confidence indicators~\cite{fomicheva-etal-2020-unsupervised,guerreiro-etal-2023-looking,kang2026scalable}. Thus, we define model confidence as the mean token log-probability of a generated response (length-normalized), yielding a sample-level logprob\_score. To enable cross-model and cross-condition comparison, this score is normalized via z-scoring within each model × condition group. The resulting normalized confidence is used to define high-confidence incorrect predictions and to compute task-level confidence~shifts.

\FloatBarrier
\section{LLM alignment with humans under code obfuscation (RQ1)}

\subsection{Accuracy under Obfuscation}
\label{acc_under_obf}

\begin{table}[htbp]
\centering
\small
\setlength{\tabcolsep}{3pt}
\caption{Obfuscation Tier vs. Accuracy (\%) for Models and Humans on Dataset A}
\label{tab:combined_obfuscation_accuracy}
\begin{tabularx}{\columnwidth}{l C C C C C C}
\hline
\textbf{Agent / Model} & \textbf{L0} & \textbf{L1} & \textbf{L1b} & \textbf{L2} & \textbf{L3} & \textbf{Max Diff} \\
\hline
% \multicolumn{7}{l}{\textit{Large Language Models}} \\
\hline
CodeLlama-7B  & 12.4 & \high{13.2} & \low{9.6}   & \low{10.8}  & \low{11.4}  & +0.8  \\
DS-Coder-6.7B & 9.4  & \high{13.4} & \high{13.4} & \high{12.4} & \high{12.2} & +4.0  \\
DS-R1-Qwen-7B & 63.8 & \high{64.2} & \low{51.8}  & \low{57.0}  & \low{56.2}  & +0.4  \\
Llama-3.1-8B  & 15.6 & \high{23.4} & \low{12.4}  & \high{16.8} & \high{19.8} & +7.8  \\
Phi-4-Mini    & 6.0  & \low{4.2}   & \low{4.4}   & \low{3.4}   & \low{3.2}   & -2.8  \\
Qwen3-0.6B    & 5.2  & 5.2         & \low{5.0}   & \high{5.8}  & \low{4.6}   & +0.6  \\
SmolLM3-3B    & 45.0 & \low{44.0}  & \low{34.2}  & \low{37.6}  & \low{33.0}  & -12.0 \\
\midrule
% \multicolumn{7}{l}{\textit{Human Participants}} \\
\hline
Human (all)   & 40.46 & \low{38.68} & \low{38.02} & \low{34.15} & \low{31.09} & \low{-9.37} \\
Beginner      & 34.21 & \low{26.67} & \low{26.32} & \low{31.43} & \low{23.08} & \low{-11.13} \\
Intermediate  & 45.10 & \low{43.14} & \low{37.50} & \low{32.76} & \low{33.33} & \low{-12.34} \\
Expert        & 40.48 & \high{44.00} & \high{55.56} & \low{40.00} & \low{37.50} & \high{+15.08} \\
\hline
\end{tabularx}
\end{table}

Table \ref{tab:combined_obfuscation_accuracy} shows performance for models and humans. Across models, accuracy generally declines as obfuscation increases (mixed-effects logistic regression, tier effect $\chi^2(4) = 45.4$, $p < 0.001$), although the pattern is rarely monotonic. Several models—including DS-Coder-6.7B, DS-R1-Qwen-7B, and Llama-3.1-8B— {\bf \em achieve their highest accuracy not on the original code (L0) but on mildly obfuscated versions (L1)}. This suggests that slight perturbations to identifiers may suppress superficial pattern matching and encourage deeper reasoning. However, this effect is limited, as accuracy typically declines at higher obfuscation tiers (L2–L3).

\begin{figure}[t]
\centering
\includegraphics[width=2.3in]{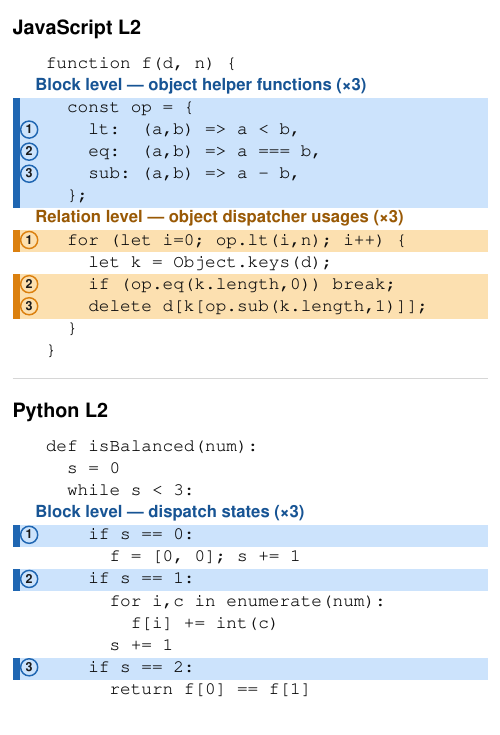}
\vspace{-15pt}
\caption{Dispatch-related metric, with structure highlighting}
\label{fig:dispatcher_example_l2_code}
\end{figure}

% Adding paragraph on dispatcher metrics (RQ4)
% --- folded in from former subsection "Why the higher tiers collapse: control-flow dispatchers" ---
This decline has a structural cause. At L2/L3, {\bf \em control-flow flattening}
imposes a new execution architecture rather than modifying existing logic:
dispatcher structures are absent through $L1b$ and appear only at L2/L3
(Fig.~\ref{fig:dispatcher_example_l2_code}). In Block Model terms, this
zero-to-nonzero transition shifts reasoning from local code units to global
execution flow, where helper functions and dispatch states act as block-level
units and dispatcher calls as relation-level indirection, obscuring macro-level
flow through implicit state tracking. Consistent with this hierarchy, dispatcher
complexity within L2/L3 correlates negatively with accuracy in both
languages---most strongly for the Python while-if state count ($r = -0.196$,
$q = 3.08\times10^{-23}$), and also for JavaScript dispatch-call references
($r$=$-0.130$, $q$=$2.76\times10^{-7}$) and object-dispatch helpers
($r$=$-0.095$, $q$=$2.69\times10^{-4}$), all significant after multiple-testing
correction. The effect is graded---state-based dispatch degrades accuracy more
than relation-level indirection, indicating that failures stem from
reconstructing execution state, not reasoning over local structure. This is not a
single-model artifact: the Python state-count effect holds across all seven
models and the JavaScript effect across~two instruction-tuned models.
Critically, the effect ties to correctness loss, not longer reasoning. Its
correlation with CoT length is insignificant ($r \approx 0.03$--$0.05$),
so the decline reflects {\bf \em genuine comprehension difficulty, not merely
longer reasoning traces}.

Performance varies considerably by architecture. {\bf \em The strongest reasoning-tuned models (DS-R1-Qwen-7B, SmolLM3-3B) outperform coders and instruction-tuned models across all tiers}, largely independent of model size. Weaker models (Phi-4-Mini, Qwen3-0.6B) show a clear floor effect: already near chance at baseline, they change little across tiers, since added obfuscation produces only small absolute shifts. Stronger models, starting from higher baselines, show larger absolute~drops.

Table~\ref{tab:side_by_side_breakdown} shows a different pattern for human participants. Overall human accuracy decreases as obfuscation increases, particularly for beginners and intermediate programmers. However, expert programmers show a different trend: {\bf \em their accuracy improves under moderate obfuscation (L1b)}, reaching the highest performance among all human groups. This suggests that experts rely less on lexical cues such as identifier names and instead reason about code structure and semantics, a
human pattern that, as shown in Section~\ref{sec:rq2}, models do not~reproduce.

\subsection{Language Effects}

Table~\ref{tab:side_by_side_breakdown} reveals a language effect that is stronger
in humans than models. Among reasoning models the Python advantage is clear only
for DS-R1-Qwen-7B (high and stable on Python while JavaScript drops at L1b);
SmolLM3-3B is weaker and inconsistent, so the effect is not a reliable main effect
but is tier-dependent (significant only at L2, $p=0.036$). Humans show a stronger,
more consistent pattern: JavaScript accuracy falls sharply with obfuscation while
Python stays robust, especially for intermediate and expert programmers, with
experts peaking on Python at L1b --- evidence of structural rather than
identifier-based reasoning. Even strong LLMs remain less robust under this
condition, highlighting a gap between model reasoning and expert human
understanding.

\begin{table}[t]
\centering
\caption{Tier Breakdown: Models vs Humans (\%) on Dataset~A} 
\label{tab:side_by_side_breakdown}
\vspace{-6pt}
\resizebox{\columnwidth}{!}{%
\setlength{\tabcolsep}{2pt}
\begin{tabular}{ll ccccccc | cccc}
\toprule
& & \multicolumn{7}{c}{\textbf{Models}} & \multicolumn{4}{c}{\textbf{Humans}} \\
\cmidrule(lr){3-9} \cmidrule(lr){10-13}
\textbf{L.} & \textbf{T.} & \makecell{CL\\7B} & \makecell{DS-C\\6.7B} & \makecell{R1\\Q7B} & \makecell{L3.1\\8B} & \makecell{Phi4\\M} & \makecell{Qw3\\0.6B} & \makecell{Smol\\3B} & \makecell{All} & \makecell{Beg.} & \makecell{Int.} & \makecell{Exp.} \\
\midrule
\textbf{JS} & L0  & 12.8 & 9.6  & 57.6 & 10.4 & 4.8  & 2.8  & 39.6 & 43.3 & 42.9 & 53.6 & 27.8 \\
            & L1  & \high{13.2} & \high{14.0} & \high{58.0} & \high{30.4} & \low{2.4}  & \high{4.0}  & \high{45.2} & \low{21.3} & \low{9.1}  & \low{20.8} & \high{33.3} \\
            & L1b & \high{14.8} & \high{15.2} & \low{38.4}  & \low{7.6}   & \low{4.0}  & \high{4.0}  & \low{26.4} & \low{26.2} & \low{28.0} & \low{20.0} & \high{36.4} \\
            & L2  & \low{9.6}   & \high{15.2} & \low{44.8}  & \high{13.6} & \low{1.6}  & \high{4.0}  & \low{26.8} & \low{25.8} & \low{29.4} & \low{21.9} & \high{30.8} \\
            & L3  & \low{11.6}  & \high{14.4} & \low{45.6}  & \high{19.2} & \low{1.6}  & \high{4.8}  & \low{30.4} & \low{15.9} & \low{9.1}  & \low{17.2} & \low{25.0} \\ 
\midrule
\textbf{Py} & L0  & 12.0 & 9.2  & 70.0 & 20.8 & 7.2  & 7.6  & 50.4 & 37.5 & 23.5 & 34.8 & 50.0 \\
            & L1  & \high{13.2} & \high{12.8} & \high{70.4} & \low{16.4}  & \low{6.0}  & \low{6.4}   & \low{42.8} & \high{52.5} & \high{36.8} & \high{63.0} & \high{53.9} \\
            & L1b & \low{4.4}   & \high{11.6} & \low{65.2}  & \low{17.2}  & \low{4.8}  & \low{6.0}   & \low{42.0} & \high{50.0} & 23.1 & \high{51.6} & \high{68.8} \\
            & L2  & 12.0        & \high{9.6}  & \low{69.2}  & \low{20.0}  & \low{5.2}  & 7.6         & \low{48.4} & \high{42.6} & \high{33.3} & \high{46.2} & \low{47.1} \\
            & L3  & \low{11.2}  & \high{10.0} & \low{66.8}  & \low{20.4}  & \low{4.8}  & \low{4.4}   & \low{35.6} & \high{48.2} & \high{41.2} & \high{57.9} & \low{45.0} \\
\bottomrule
\end{tabular}}
\vspace{3pt}
\end{table}

\subsection{Human--model Alignment}

\begin{figure}[t]
    \centering
    \includegraphics[width=\linewidth]{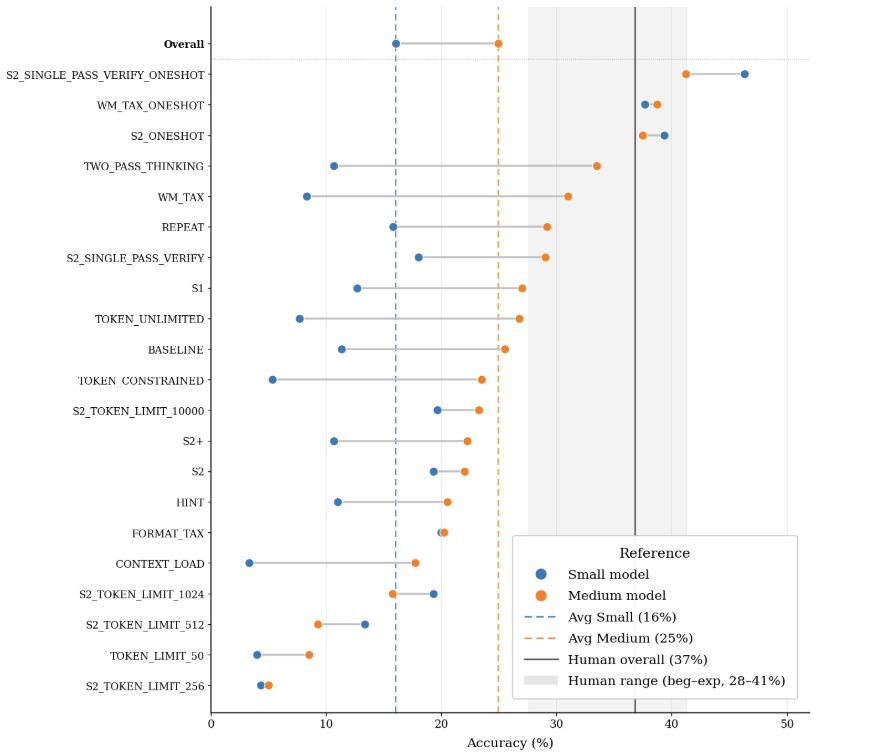}
    \vspace{-18pt}
    \caption{Accuracy across conditions for reasoning, instruct, and coder models.}
    \label{fig:conditions}
\end{figure}

The strongest performers are {\bf \em DeepSeek-R1-Qwen-7b and SmolLM3-3b}, both of which demonstrated robust reasoning capabilities. Conditional on producing a final answer, DS-R1-Qwen-7B is correct 58.8\% of the time, while SmolLM3-3B is correct 49.6\% of the time. Fig.~\ref{fig:conditions} shows that {\bf \em the strongest reasoning models (DS-R1-Qwen-7B, and marginally SmolLM3-3B) exceed human baselines but collapse under low token limits} (OR=12.5, 95\% CI [6.4,24.2], p<0.001). This suggests that their performance is a function of System 2. All models show a performance lift with \code{S2 + One-Shot} prompts. This suggests that combining explicit reasoning paths with in-context examples effectively reduces the logical search space. Finally, the negligible impact of "WM Tax" suggests that, unlike humans, model failures are driven by logical execution bottlenecks rather than information retention limits.

\begin{table}[t]
\caption{Correlation between human accuracy and model accuracy. Significant correlations ($p < 0.05$) are highlighted.}
\centering
\small
\setlength{\tabcolsep}{4pt}
\begin{tabular}{l l l c c}
\hline
\textbf{Model} & \textbf{Size} & \textbf{Type} & \textbf{Spearman $\rho$} & \textbf{$p$-value} \\
\hline
Qwen3-0.6B & Small & Reasoning & \high{0.30} & \high{0.003} \\
Phi-4 Mini (4B) & Small & Reasoning & \high{0.47} & \high{$< 0.001$} \\
SmolLM3-3B & Small & Reasoning & \high{0.36} & \high{$< 0.001$} \\
DS-R1 Qwen-7B & Medium & Reasoning & \high{0.37} & \high{$< 0.001$} \\
\hline
Llama 3.1-8B & Medium & Instruct & 0.08 & 0.440 \\
CodeLlama 7B & Medium & Coder & 0.02 & 0.838 \\
DS-Coder 6.7B & Medium & Coder & 0.10 & 0.349 \\
\hline
\end{tabular}
\label{tab:rq1_human_model_corr}
\end{table}

Table \ref{tab:rq1_human_model_corr} assesses if model difficulty aligns with human performance at the task level. {\bf \em Reasoning models correlate significantly with human accuracy; coder and instruct models do not.} Among reasoning models, Phi-4 Mini has the highest correlation ($\rho = 0.47$), followed by DS-R1-Qwen-7B, SmolLM3-3B, and Qwen3-0.6B. CodeLlama-7B and DS-Coder-6.7B show near-zero correlations.

Notably, the {\bf \em alignment and raw accuracy are distinct}: DS-R1-Qwen-7B is the strongest overall performer, but Phi-4 Mini best mirrors human task-level behavior. Similarly, positive correlation doesn't imply high performance, even weak reasoning models correlate with humans, suggesting they share the same sources of difficulty even when failing.

\FloatBarrier

\begin{tcolorbox}[colback=white, colframe=black, arc=8pt, boxrule=0.5pt]
\textbf{RQ1 Takeaway:} \textit{Reasoning-tuned models show
significant alignment with human task-level difficulty
(Spearman $\rho = 0.30$--$0.47$), tracking which obfuscated tasks
humans find hard. Coder and instruct-tuned models show near-zero correlation.}
\end{tcolorbox}
\section{How is model alignment with human experience? (RQ2)}
\label{sec:rq2}

\begin{table}[!htbp]
\centering
\scriptsize
\setlength{\tabcolsep}{4.5pt}
\caption{Accuracy vs. obfuscation level by model type}
\label{tab:acc_vs_obf_level_complete}
\vspace{-6pt}
\begin{tabular}{lll ccccc}
\toprule
\textbf{Model} & \textbf{Size} & \textbf{Type} & \textbf{L0} & \textbf{L1} & \textbf{L1b} & \textbf{L2} & \textbf{L3} \\
\midrule
CodeLlama 7B    & Medium & Coder     & \acc{12} & \acc{13} & \acc{10} & \acc{11} & \acc{11} \\
DS-Coder 6.7B   & Medium & Coder     & \acc{9}  & \acc{13} & \acc{13} & \acc{12} & \acc{12} \\
\midrule
Llama 3.1-8B    & Medium & Instruct  & \acc{16} & \acc{23} & \acc{12} & \acc{17} & \acc{20} \\
\midrule
Qwen3-0.6B      & Small  & Reasoning & \acc{5}  & \acc{5}  & \acc{5}  & \acc{6}  & \acc{5}  \\
Phi-4 Mini (4B) & Small  & Reasoning & \acc{6}  & \acc{4}  & \acc{4}  & \acc{3}  & \acc{3}  \\
SmolLM3-3B      & Small  & Reasoning & \acc{45} & \acc{44} & \acc{34} & \acc{38} & \acc{33} \\
DS-R1 Qwen-7B   & Medium & Reasoning & \acc{64} & \acc{64} & \acc{52} & \acc{57} & \acc{56} \\
\bottomrule
\end{tabular}
\end{table}

As seen in Table \ref{tab:acc_vs_obf_level_complete}, reasoning model performance scales with size. DS-R1 Qwen-7B leads all models at 64\% baseline accuracy, with the notable exception of Phi-4 Mini, which underperforms despite its 4B parameters. Coder and instruct models occupy the middle of the range, above the floor-bound small reasoning models but well below DS-R1-Qwen-7B and SmolLM3-3B. Across all models, accuracy declines across obfuscation tiers (no model shows a strictly monotonic trend).

We divide prompts into two broad categories: System~1 (quick thinking, e.g., prompting not to think) and System~2 (deliberate thinking, e.g., prompting to take time or double-check), comprising 14 and 7 prompts respectively. 

\begin{figure}[t] %[htbp]
  \centering
  \includegraphics[width=\columnwidth]{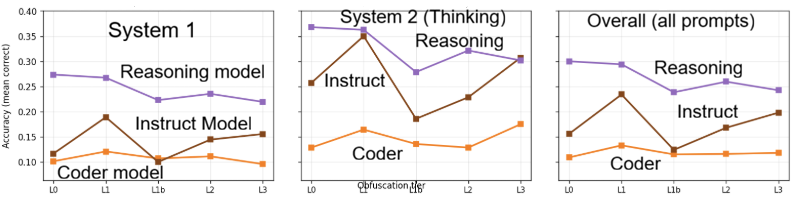}
  \vspace{-15pt}
  \caption{Accuracy vs. obfuscation tier (L0--L3) by model size and prompt setting.}
  \label{fig:size}

  \vspace{6pt}
  \includegraphics[width=\columnwidth]{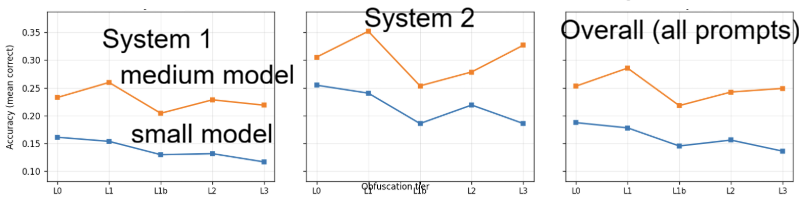}
  \vspace{-15pt}
  \caption{Accuracy vs. obfuscation tier (L0--L3) by model type and prompt setting.}
  \label{fig:type}
\end{figure}

As shown in Fig.~\ref{fig:size}, medium models outperform small models at every obfuscation tier across all prompt conditions, though System~2 prompting induces a slight accuracy dip at L1b. Prompt condition modulates absolute performance but does not alter the relative size ordering.

Reasoning models consistently outperform coder and instruct models across all prompt conditions and obfuscation tiers, with the gap most clearly under System~2 prompting (Fig.~\ref{fig:type}). Instruct and coder models perform comparably, though instruct models exhibit a notable sensitivity peak at L1b under System~2, suggesting heightened reliance on semantic cues, before recovering at higher tiers. Reasoning models are the most stable across obfuscation tiers, though not strictly monotonic.

\begin{table}[!htbp]
\centering
\caption{Accuracy by obfuscation tier and model on Dataset A divided by language}
\label{tab:acc_js_py_compact}
\small
\setlength{\tabcolsep}{6pt} 
\vspace{-6pt}
\resizebox{\columnwidth}{!}{%
\begin{tabular}{l ccccc}
\toprule
\textbf{Model} & \textbf{L0} & \textbf{L1} & \textbf{L1b} & \textbf{L2} & \textbf{L3} \\
\textit{Format: JS / PY} & \textit{acc} & \textit{acc} & \textit{acc} & \textit{acc} & \textit{acc} \\
\midrule
Qwen3-0.6B (Small)      & 3\% / 8\%   & 4\% / 6\%   & 4\% / 6\%   & 4\% / 8\%   & 5\% / 4\%   \\
Phi-4 Mini (4B) (Small) & 5\% / 7\%   & 2\% / 6\%   & 4\% / 5\%   & 2\% / 5\%   & 2\% / 5\%   \\
SmolLM3-3B (Small)      & 40\% / 50\% & 45\% / 43\% & 26\% / 42\% & 27\% / 48\% & 30\% / 36\% \\
\midrule
DS-R1 Qwen-7B (Medium)  & 58\% / 70\% & 58\% / 70\% & 38\% / 65\% & 45\% / 69\% & 46\% / 67\% \\
Llama 3.1-8B (Medium)   & 10\% / 21\% & 30\% / 16\% & 8\% / 17\%  & 14\% / 20\% & 19\% / 20\% \\
CodeLlama 7B (Medium)   & 13\% / 12\% & 13\% / 13\% & 15\% / 4\%  & 10\% / 12\% & 12\% / 11\% \\
DS-Coder 6.7B (Medium)  & 10\% / 9\%  & 14\% / 13\% & 15\% / 12\% & 15\% / 10\% & 14\% / 10\% \\
\bottomrule
\end{tabular}%
}
\end{table}

Table~\ref{tab:acc_js_py_compact} shows that the size-dependent performance pattern holds within each language. DS-R1 Qwen-7B is the top performer in both JavaScript and Python, reaching 58\% and 70\% at L0 respectively. SmolLM3-3B stands out among small models, achieving comparable baseline accuracy to some medium models (JS: 40\%, PY: 50\%), though it degrades more steeply under obfuscation.

\begin{table}[t]
\scriptsize
\centering
\caption{Accuracy by obfuscation tier and model on Dataset A. Each cell reports Baseline/S2\_ONESHOT prompts (for details on prompts, see Section~\ref{subsec:prompts}) accuracy.}
\label{tab:acc_baseline_s2_compact}
\vspace{-6pt}
\tabcolsep 4.3pt
{\scriptsize
\begin{tabular}{lccccc}
\toprule
\textbf{Model} & \textbf{L0} & \textbf{L1} & \textbf{L1b} & \textbf{L2} & \textbf{L3} \\
\midrule
Qwen3-0.6B (Small) & \pair{0}{25}\% & \pair{0}{30}\% & \pair{0}{15}\% & \pair{0}{45}\% & \pair{0}{50}\% \\

Phi-4 Mini (4B) (Small) & \pair{0}{65}\% & \pair{0}{35}\% & \pair{0}{45}\% & \pair{0}{15}\% & \pair{0}{45}\% \\

SmolLM3-3B (Small) & \pair{25}{55}\% & \pair{35}{55}\% & \pair{25}{25}\% & \pair{45}{45}\% & \pair{40}{40}\% \\

\midrule

DS-R1 Qwen-7B (Medium) & \pair{70}{65}\% & \pair{65}{70}\% & \pair{55}{45}\% & \pair{70}{55}\% & \pair{60}{80}\% \\

Llama 3.1-8B (Medium) & \pair{5}{40}\% & \pair{10}{45}\% & \pair{15}{35}\% & \pair{10}{25}\% & \pair{30}{35}\% \\

CodeLlama 7B (Medium) & \pair{30}{35}\% & \pair{25}{30}\% & \pair{20}{25}\% & \pair{15}{15}\% & \pair{25}{35}\% \\

DS-Coder 6.7B (Medium) & \pair{0}{20}\% & \pair{0}{25}\% & \pair{0}{30}\% & \pair{0}{20}\% & \pair{5}{20}\% \\

\bottomrule
\end{tabular}
}
\end{table}

Coder models consistently underperform reasoning and instruct models at the same size across both languages, and Python baseline accuracy is generally higher than JavaScript, though relative model rankings remain stable between languages. Obfuscation sensitivity is greater in JavaScript, particularly at L1b, suggesting language-specific features amplify difficulty without altering capability rankings. The \code{S2\_ONESHOT} prompt (Table~\ref{tab:acc_baseline_s2_compact}) benefits models with near-zero baseline accuracy most strongly (Qwen3-0.6B and Phi-4 Mini improve from 0\% to non-trivial accuracy across most tiers), while SmolLM3-3B benefits least. DS-R1 Qwen-7B shows a mixed pattern: gains at L1 and L3, slight reductions at L0 and L1b.

\begin{figure}[t]
    \centering
    \includegraphics[width=1\linewidth]{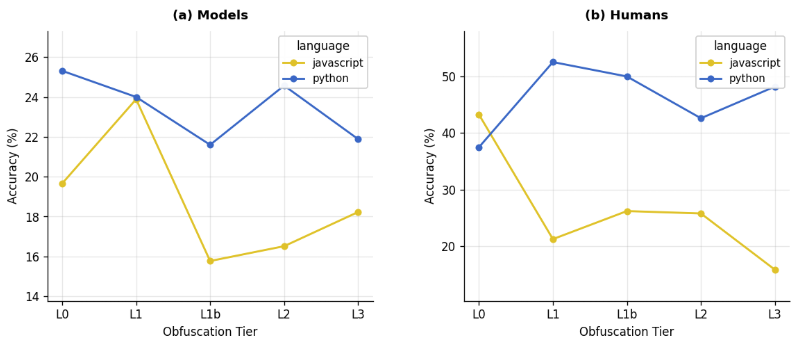}
    \vspace{-12pt}
    \caption{Accuracy (\%) by programming language (JavaScript vs. Python) across obfuscation tiers (L0–L3). Left: model performance averaged across all models. Right: human performance across all experience levels, reproduced from ~\cite{nguyen2026effectcodeobfuscationhuman}.
}
    \label{fig:langu}
\end{figure}

Fig.~\ref{fig:langu} (right) shows that human performance varies substantially across languages, with consistently higher accuracy in Python compared to JavaScript across all obfuscation tiers (L1-L3) \cite{nguyen2026effectcodeobfuscationhuman}. In contrast, Fig.~\ref{fig:langu} (left) illustrates that while models also exhibit language-dependent variation, these differences are less pronounced and follow a different pattern of degradation.
If models align with human expertise, we would expect similar relative trends across language and tiers. However, the divergence between human and model performance patterns suggests that models do not possess the same language-dependent comprehension behavior observed in humans. These differences suggest model performance may be influenced more by factors such as capacity and architectural specialization than by language-specific experience.

\begin{figure}[t]
    \centering
    \includegraphics[width=\linewidth]{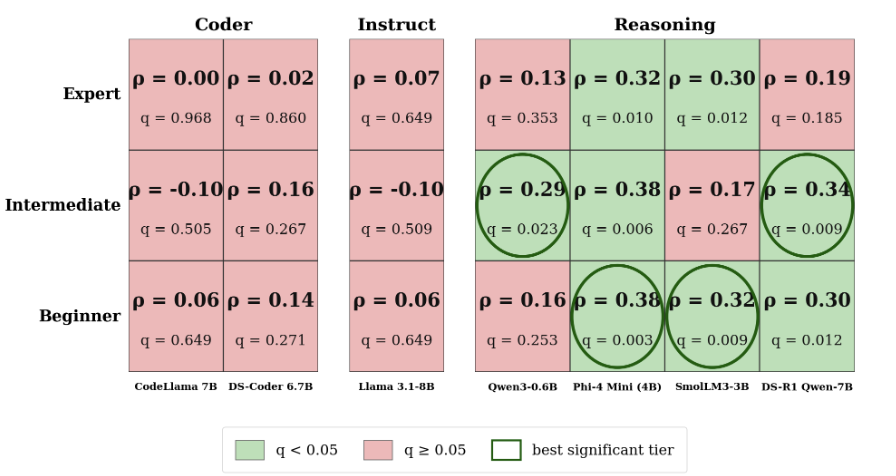}
    \vspace{-15pt}
    \caption{Human–model alignment by experience level. Each cell
    reports Spearman $\rho$ and BH-FDR $q$; bordered cells are
    significant ($q<.05$). Reasoning models show significant positive
    alignment across all three tiers, with no tier reliably
    distinguishable from the others.}
    \label{fig:exp_corr}
\end{figure}

To directly assess alignment with human expertise levels, we compare
model performance patterns with human groups stratified by experience.
Fig.~\ref{fig:exp_corr} shows that reasoning models exhibit significant
positive correlations across all three experience tiers, with no tier
reliably distinguishable from the others (mean $\rho \approx 0.24$--$0.30$;
leave-one-model-out ordering is unstable and a mixed model finds no
tier~$\times$~alignment interaction). Coder and instruct models, by
contrast, show statistically insignificant correlations across all
experience levels. The presence of alignment across every tier
exclusively in reasoning models reinforces the RQ1 finding: reasoning
models succeed and fail on the same tasks as human participants and share
their difficulty profiles across obfuscation tiers, whereas coder and
instruct models show no such alignment, implying their failures are driven
by fundamentally different processes. Reasoning-tuned models approximate the
human difficulty patterns that obfuscation induces, while coder and
instruct models do not~\cite{nguyen2026effectcodeobfuscationhuman}.

\begin{tcolorbox}[colback=white, colframe=black, arc=8pt, boxrule=0.5pt]
\textbf{RQ2 Takeaway:} The strongest reasoning models outperform coder and
instruct models and align with human difficulty patterns across
all experience tiers, with no tier robustly distinguishable from
the others. No model reproduces the expert accuracy
inversion under~L1b.
\end{tcolorbox}

\section{Chain of Thought (CoT) Trace Lengths (RQ3)}

\begin{figure}[t]
    \centering
    \includegraphics[width=0.65\linewidth]{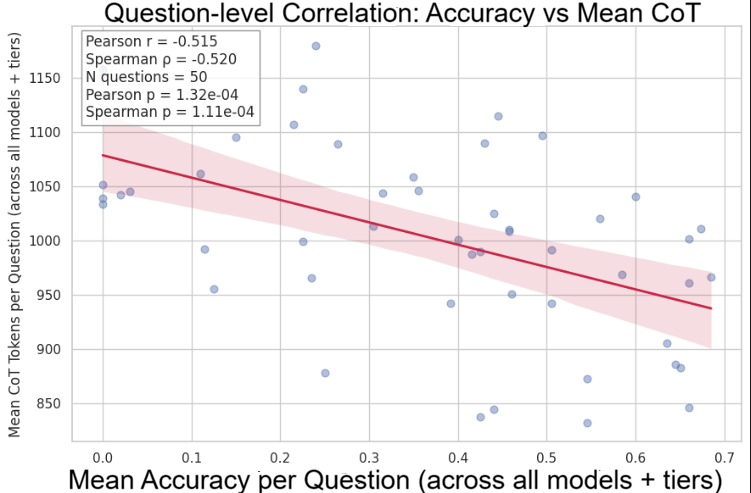}
    \caption{Mean Accuracy per Question vs. Mean CoT Tokens per Answer  with 95\% confidence thresholds}

    \label{fig:accuracy_mean}

\end{figure}

Unlike RQ1–RQ2, which rely on Dataset A for human–model alignment, the analyses in this section are computed over Dataset B, our larger model-only output-prediction set spanning all five obfuscation tiers.

Fig.~\ref{fig:accuracy_mean} shows a significant negative correlation between mean accuracy and mean CoT trace length (Spearman $\rho = -0.520$, $p = 1.11 \times 10^{-4}$; Pearson $r = -0.515$, $p = 1.32 \times 10^{-4}$), persistent across models, prompts, and obfuscation tiers.

As shown in Table~\ref{fig:strat_cot}, incorrect predictions have descriptively
higher mean token counts than correct ones, exceeding 1{,}000 tokens across
L1--L3, though this difference is not significant once snippet and model
are accounted for ($p=0.33$). Mean token usage for incorrect answers is also
nearly identical between structural (L1: 1{,}011.4) and semantic (L1b: 1{,}014.4)
obfuscation, indicating that added obfuscation does not translate into
proportionally longer reasoning on failed attempts.

In contrast, successful predictions show a monotonic increase in
token length from L0 to L2 (negative-binomial regression, $p=0.004$),
indicating that models expend progressively more effort on
increasingly obfuscated code. Correct-response length does not differ
between L1 and L1b (negative-binomial GLMM, estimated marginal means
contrast, ratio $=1.01$, 95\% CI $[0.91,1.12]$, $p=0.85$). The negative
accuracy--length correlation across questions is driven by task
difficulty: harder questions elicit longer traces and are also
answered incorrectly more often.

A notable exception occurs at L3, where correct responses average fewer tokens than at L2, likely survivorship bias:~compounding complexity leaves only the simplest programs (e.g., trivial boolean
returns) solvable, which require shorter traces. 

\begin{table}[t]
\caption{Overall mean CoT trace length stratified by correct/incorrect and obfuscation tier}
\small
%\vspace{-3pt}
\begin{tabular}{lccccc}
\toprule
\textbf{Outcome} & \multicolumn{5}{c}{\textbf{Obfuscation Tier}} \\
\cmidrule(lr){2-6}
 & L0 & L1 & L1b & L2 & L3 \\
\midrule
Correct & \cellcolor[HTML]{FFFFCC} 910.7 & \cellcolor[HTML]{FFEEA3} 924.9 & \cellcolor[HTML]{FEA044} 962.1 & \cellcolor[HTML]{FD7034} 976.3 & \cellcolor[HTML]{FEDC7C} 938.1 \\
Incorrect & \cellcolor[HTML]{FC6832} 978.1 & \cellcolor[HTML]{C20325} \textcolor{white}{1011.4} & \cellcolor[HTML]{B90026} \textcolor{white}{1014.4} & \cellcolor[HTML]{8F0026} \textcolor{white}{1024.3} & \cellcolor[HTML]{800026} \textcolor{white}{1028.3} \\
\bottomrule
\end{tabular}
\label{fig:strat_cot}
\end{table}

\begin{figure}[htbp]
  \centering
  \begin{minipage}{0.48\columnwidth}
    \centering
    \includegraphics[width=\linewidth]{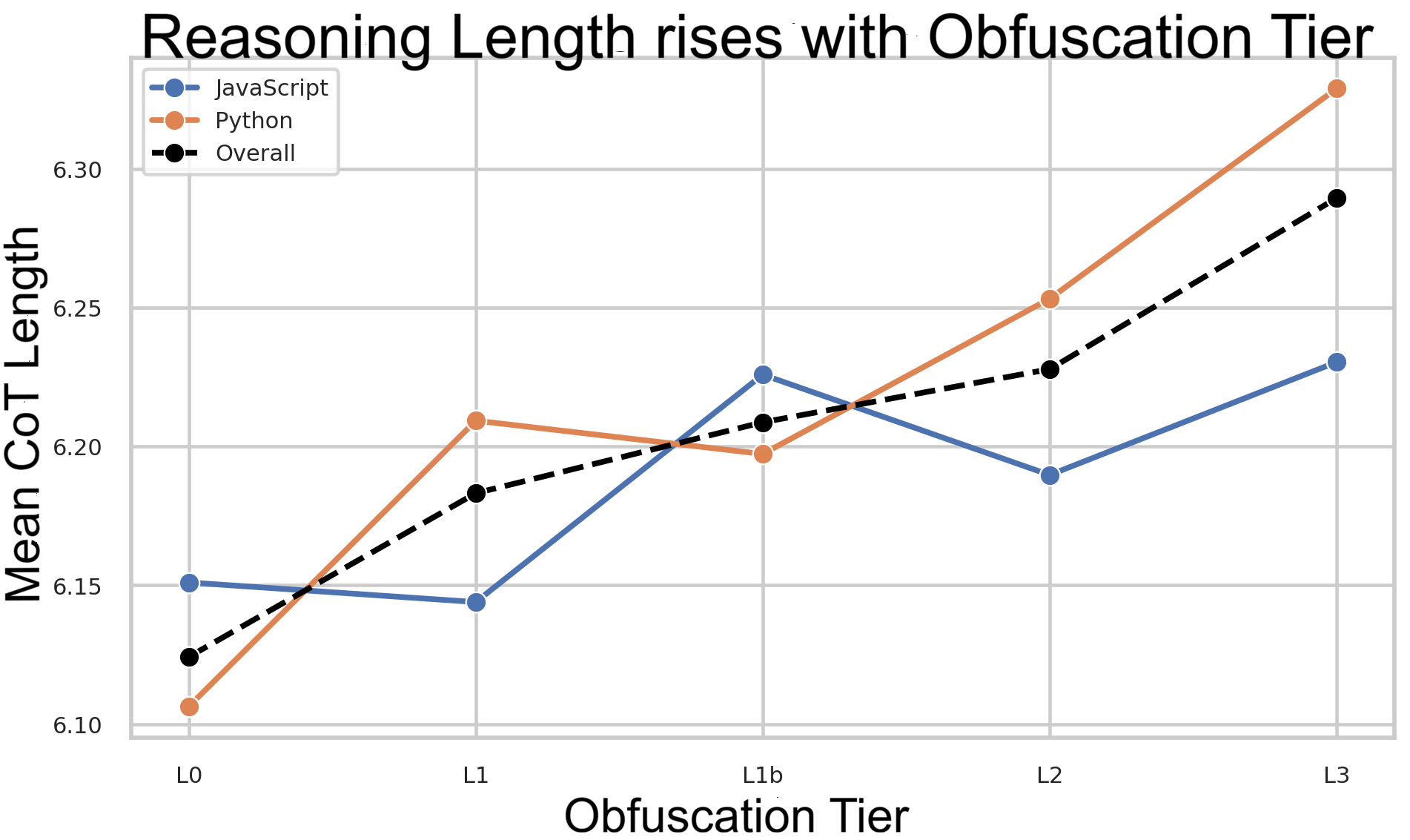}
    \vspace{-15pt}
    \caption{Log length of CoT}
    \label{fig:log-cot}
  \end{minipage}% <--- The % is critical to prevent a line break
  \hfill
  \begin{minipage}{0.48\columnwidth}
    \centering
    \includegraphics[width=\linewidth]{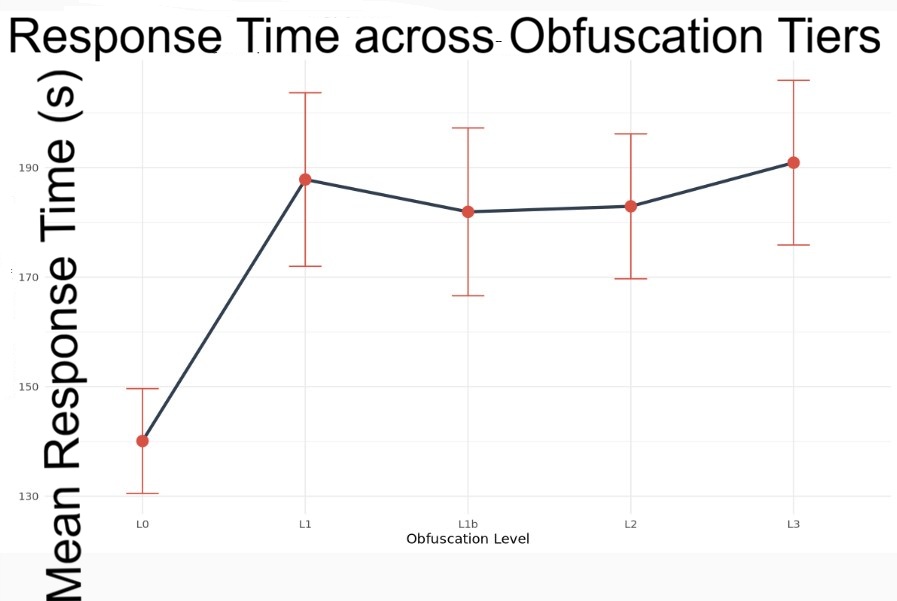}
    \vspace{-15pt}
    \caption{Human response time~\cite{nguyen2026effectcodeobfuscationhuman}}
    \label{fig:human-response-time}
  \end{minipage}
\end{figure}

Figs.~\ref{fig:log-cot} and \ref{fig:human-response-time} contrast how models and humans respond to increasing code obfuscation. For models, higher obfuscation generally drives a monotonic, language-agnostic increase in mean Chain-of-Thought (CoT) length, culminating in a significant spike at L3. The only exception is L1b, which exhibits minor language-specific variances. In contrast to this steady scaling of model effort, human response times jump significantly from L0 to L1 but remain flat through L3. This divergence suggests that {\bf while humans utilize distinct cognitive processing modes (System 1 and System 2), models scale their computational effort more monotonically}.

\begin{figure}[t]
    \centering
    \includegraphics[width=0.8\linewidth]{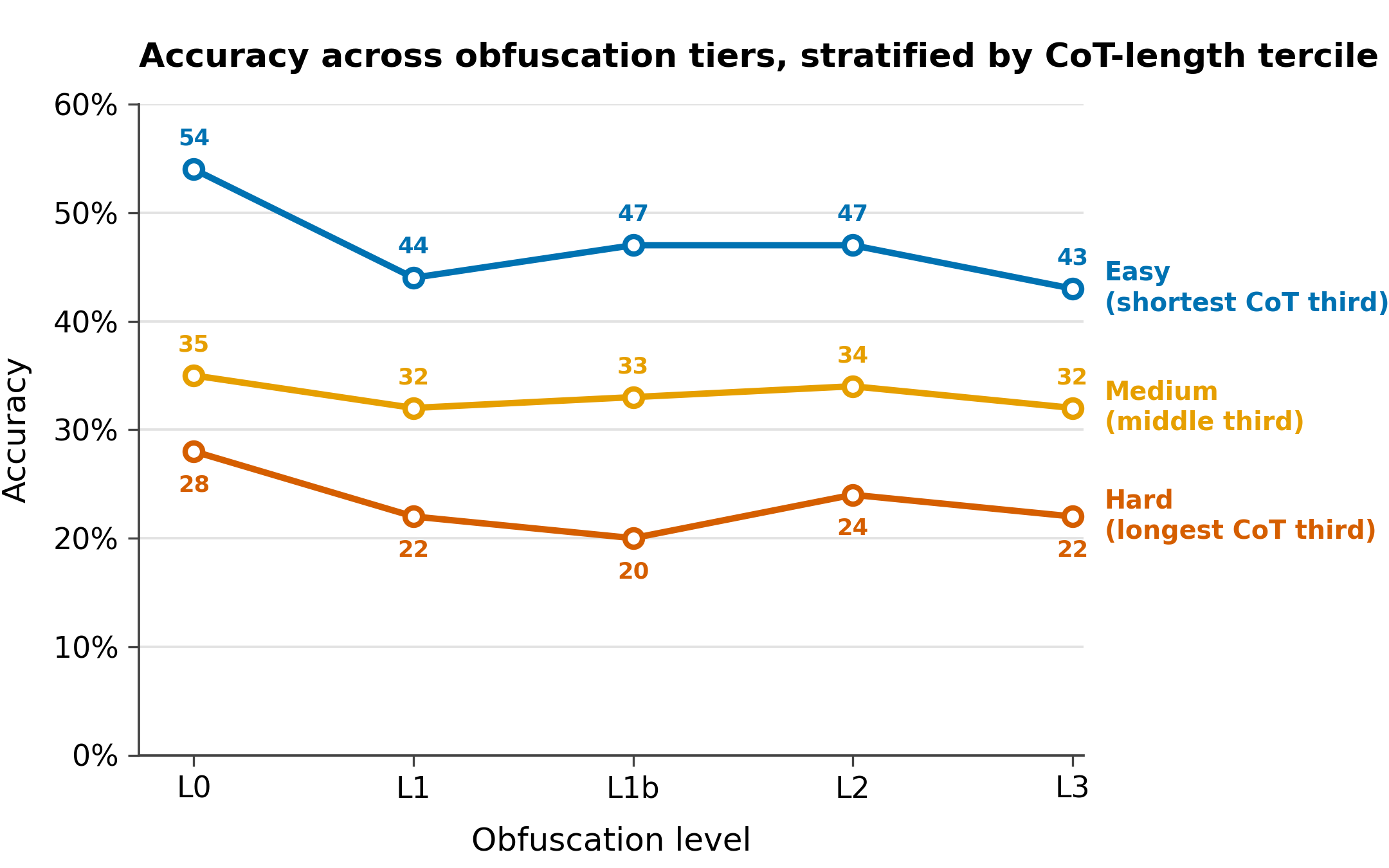}
    \vspace{-3pt}
    \caption{Accuracy by obfuscation tiers (L0--L3), stratified by CoT length tercile (easy = shortest third, hard = longest third).}
    \label{fig:bin}
\end{figure}

Fig.~\ref{fig:bin} stratifies tasks by Chain-of-Thought length tercile: the shortest third of traces are labeled easy, the longest third hard. The three terciles stay cleanly separated and roughly flat across all tiers, with easy tasks near 43--54\% and hard tasks around 20--28\%. This persistent separation from L0 to L3 reinforces the RQ3 finding: the tasks eliciting the longest traces are consistently those the model gets wrong, indicating that trace length tracks with difficulty.

\begin{tcolorbox}[colback=white, colframe=black, arc=8pt, boxrule=0.5pt]
\textbf{RQ3 Takeaway:} \textit{CoT length tracks task difficulty
rather than serving as a confusion signal: the tasks eliciting the
longest traces are consistently those the model answers incorrectly,
so length marks difficulty across questions ($\rho = -0.52$). CoT
length also rises with obfuscation tier ($p = 0.004$).}
\end{tcolorbox}

\section{Adversarial renaming (RQ4)}
\label{sec:rq4}

Having established that reasoning models track which tasks humans find hard (RQ1–RQ2), we now ask whether they also reproduce a specific human failure mode: confident misinterpretation under misleading identifiers.

Prior work shows identifier names can mislead neural models 
under semantics-preserving transformations~\cite{yefet2020adversarialexamples, 
wainakh2021idbench}, but does not quantify the semantic 
displacement introduced or its effect on confidence--accuracy 
relationships and high-confidence errors. 
Building on this, we situate adversarial renaming at the 
\textit{atom level} of the Block Model, isolating whether 
comprehension failures arise from lexical-semantic 
misinterpretation rather than structural complexity.

\begin{figure}[t]
\centering
\includegraphics[width=2.2in] %0.45\linewidth
{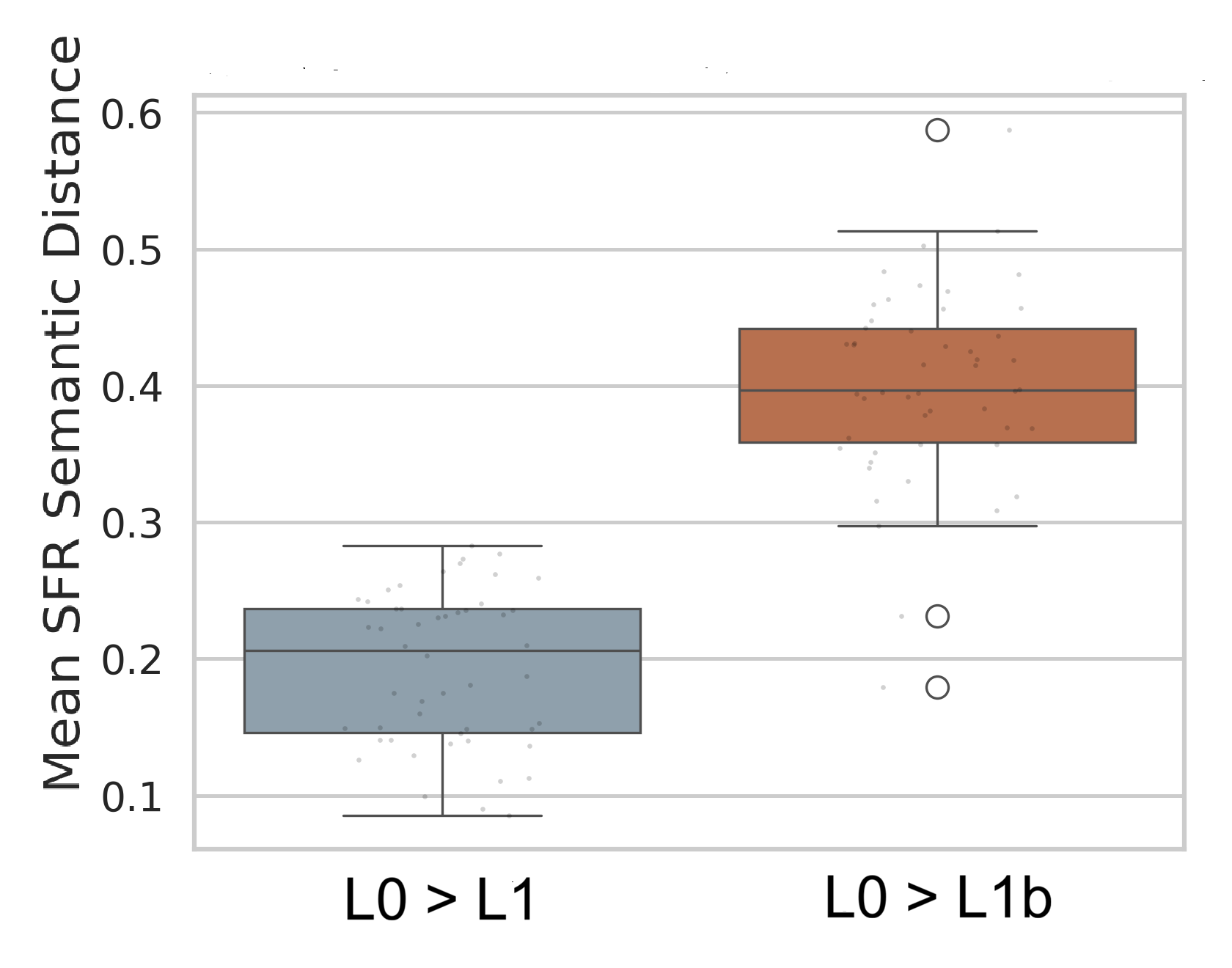}
\vspace{-6pt}
\caption{Distribution of semantic distance (SFR embeddings) for L0→L1 and L0→L1b. Points show snippet-level means; boxplots summarize distributions.}
\label{fig:valid_semantic_displacement}
\end{figure}

Fig.~\ref{fig:valid_semantic_displacement} confirms construct validity: $L1b$ produces substantially higher semantic distance than L1 (standard renaming), with well-separated distributions and minimal overlap.

\begin{figure}[t]%[!htbp]
    \centering
    \begin{minipage}[t]{0.5\linewidth}
        \centering
        \textbf{(a) Tercile stratification}\par\vspace{2pt}
        \includegraphics[width=\linewidth]{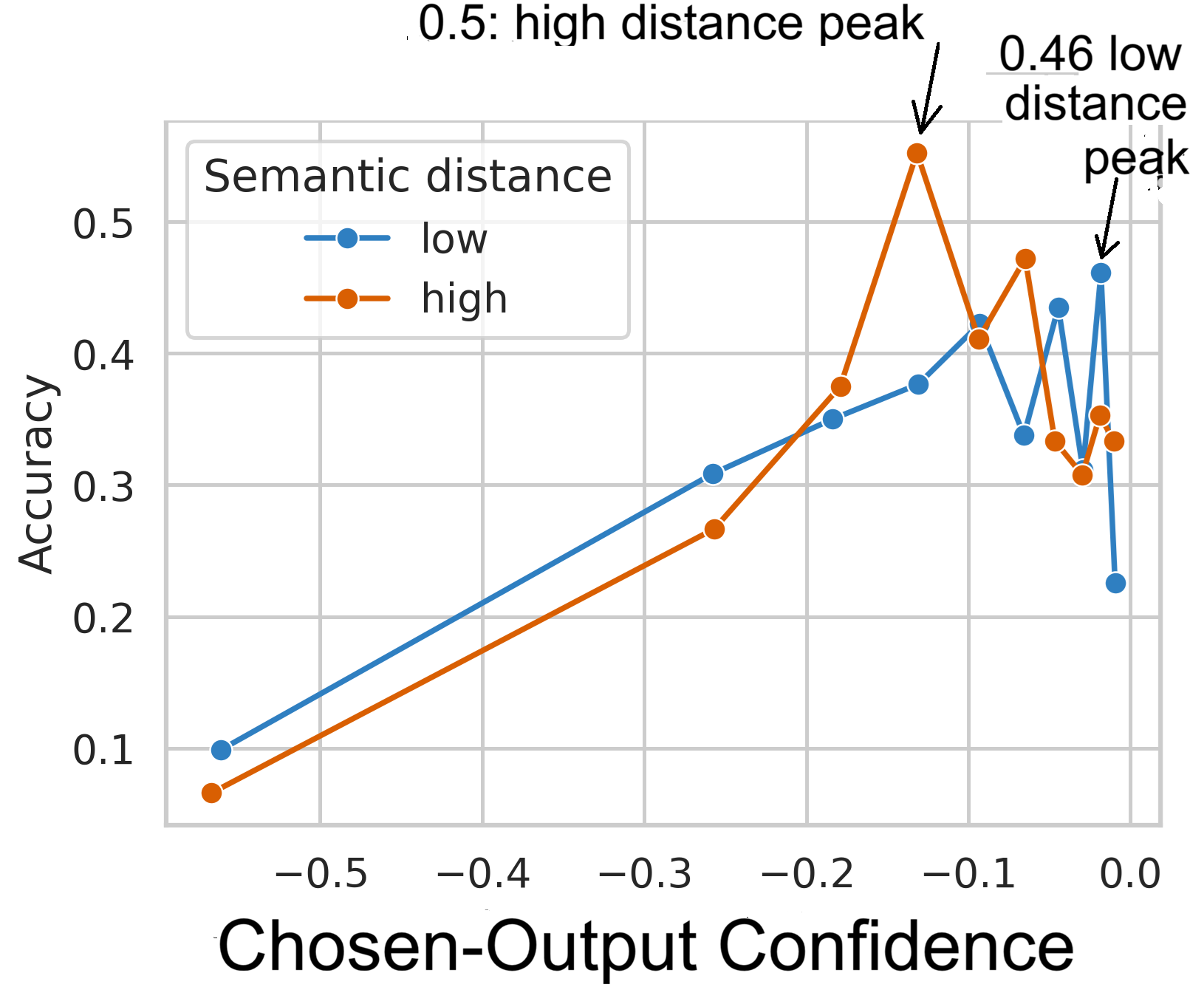}
    \end{minipage}\hfill
    \begin{minipage}[t]{0.49\linewidth}
        \centering
        \textbf{(b) Decile stratification}\par\vspace{2pt}
        \includegraphics[height=1.2in]{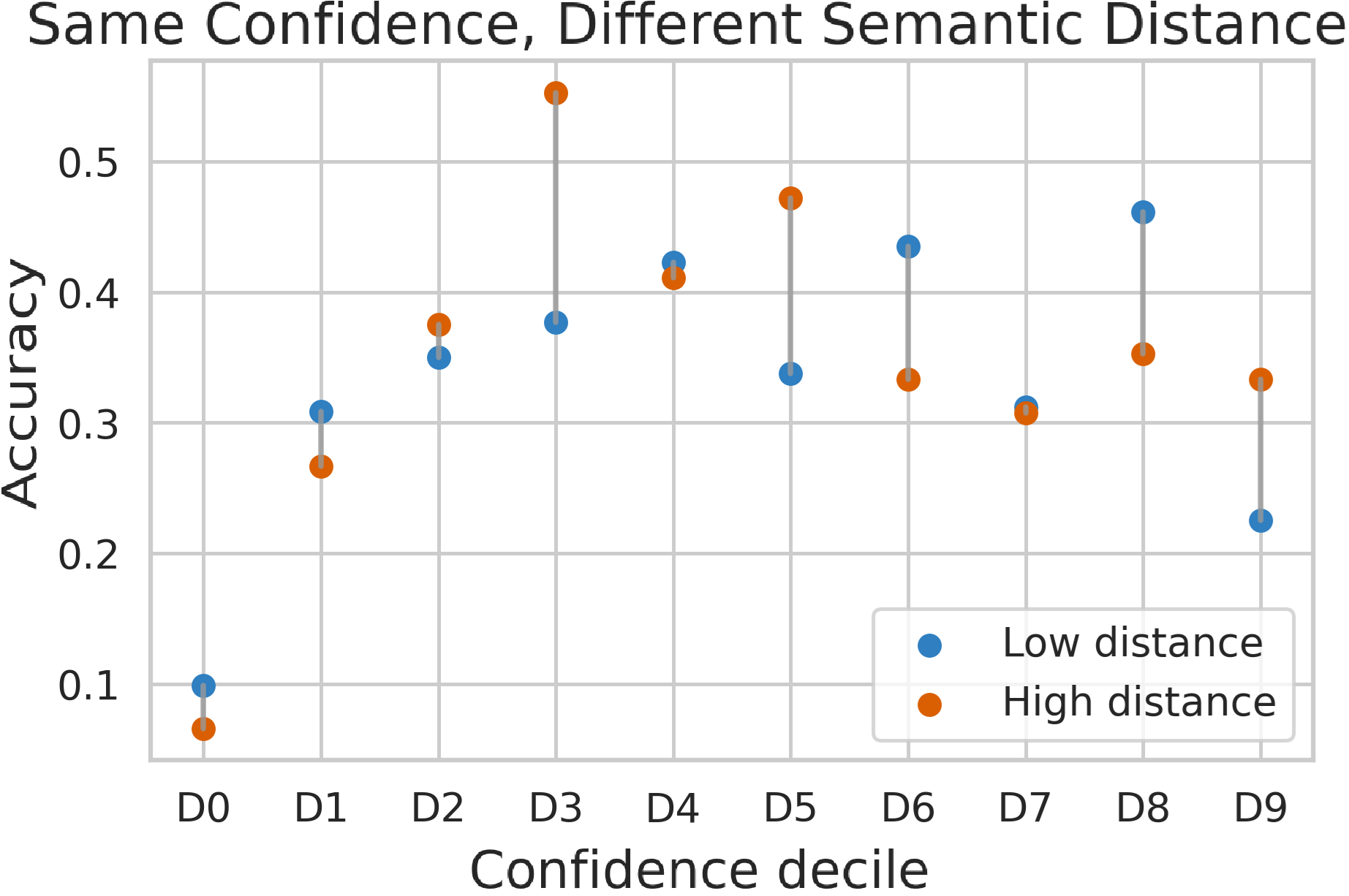}
    \end{minipage}
    \vspace{-3pt}
     \caption{Confidence--accuracy relationship in L1b across semantic-displacement strata. Left: terciles; Right: deciles.}
    \label{fig:confidence_acc_distance_combined}
\end{figure}

Fig.~\ref{fig:confidence_acc_distance_combined} examines model behavior within $L1b$, stratified by semantic mis-framing magnitude ($D_{L0 \rightarrow L1b} - D_{L0 \rightarrow L1}$). In the low-confidence region, high- and low-displacement subsets behave similarly. In the mid-confidence range, however, they diverge: the low-displacement subset shows a gradual, near-monotonic increase in accuracy ($\sim$0.38--0.42), consistent with a well-calibrated confidence--accuracy relationship, whereas the high-displacement subset exhibits non-monotonic behavior, peaking at the third decile ($\sim$0.55) before dropping sharply ($\sim$0.33) at higher confidence levels. At the highest confidence decile, both subsets collapse, likely due to degenerate outputs in which the model enters a repetitive generation loop and assigns near-certain probability to an incorrect token sequence.

Reorganizing the same data by matched confidence deciles (Fig.~\ref{fig:confidence_acc_distance_combined}b) shows that the gap between subsets varies inconsistently across deciles, with no stable ordering. This indicates that semantic displacement does not systematically shift accuracy at a fixed confidence level, but instead destabilizes the confidence--accuracy relationship: predictions at similar confidence can yield substantially different outcomes depending on displacement magnitude. From a Block Model perspective, this instability originates at the \textit{atom level} --- misleading identifier semantics disrupt the mapping between surface cues and underlying computation without altering structural reasoning. This variability motivates a finer-grained analysis of where confident errors concentrate, which we examine through HCI and identifier-level disruption below.

\begin{table}[t]
\caption{Localization results by semantic distance and identifier-spike regime. $\Delta$ measures the change from low spike to medium spike within each semantic-distance band.}
\centering
\small
\setlength{\tabcolsep}{4pt}
%\vspace{-3pt}
\begin{tabular}{lccc|ccc}
\hline
& \multicolumn{3}{c|}{\textbf{Accuracy}} & \multicolumn{3}{c}{\textbf{HCI}} \\
\textbf{Sem. Dist.} 
& \makecell{\textbf{Low}\\\textbf{Spike}} 
& \makecell{\textbf{Med.}\\\textbf{Spike}} 
& $\Delta$ 
& \makecell{\textbf{Low}\\\textbf{Spike}} 
& \makecell{\textbf{Med.}\\\textbf{Spike}} 
& $\Delta$ \\
\hline
Low    
& 29.44\% & 38.61\% & \high{+9.18\%}  
& 1.52\% & 5.94\%  & \high{+4.42\%} \\

Medium 
& 32.51\% & 30.96\% & \low{-1.55\%}   
& 3.99\% & 8.79\%  & \high{+4.80\%} \\

High   
& 36.33\% & 31.25\% & \low{-5.08\%}   
& 5.66\% & 5.77\%  & \high{+0.11\%} \\
\hline
\end{tabular}
\label{tab:figure3_localization_transition}
\end{table}

\begin{table}[t]
\caption{Accuracy and HCI rate across L1b tail-regime subsets. Low/Low baseline = both metrics in bottom 20\%.}
\centering
\small
%\vspace{-3pt}
\begin{tabular}{lrrr}
\hline
\textbf{Regime} & \textbf{n} & \textbf{Accuracy} & \textbf{HCI Rate} \\
\hline
Low/Low baseline      & 202 & 31.68\% & 4.95\% \\
Top 20\% distance     & 450 & \high{33.11\%} & \high{6.89\%} \\
Top 20\% spike        & 450 & \high{36.00\%} & \high{8.44\%} \\
Top distance $\cap$ spike & 80  & \low{21.25\%} & \high{10.00\%} \\
\hline
\end{tabular}
\label{tab:spike_distance_subsets}
\end{table}

\begin{figure}[t]
    \centering
    \includegraphics[width=0.72\linewidth]
    {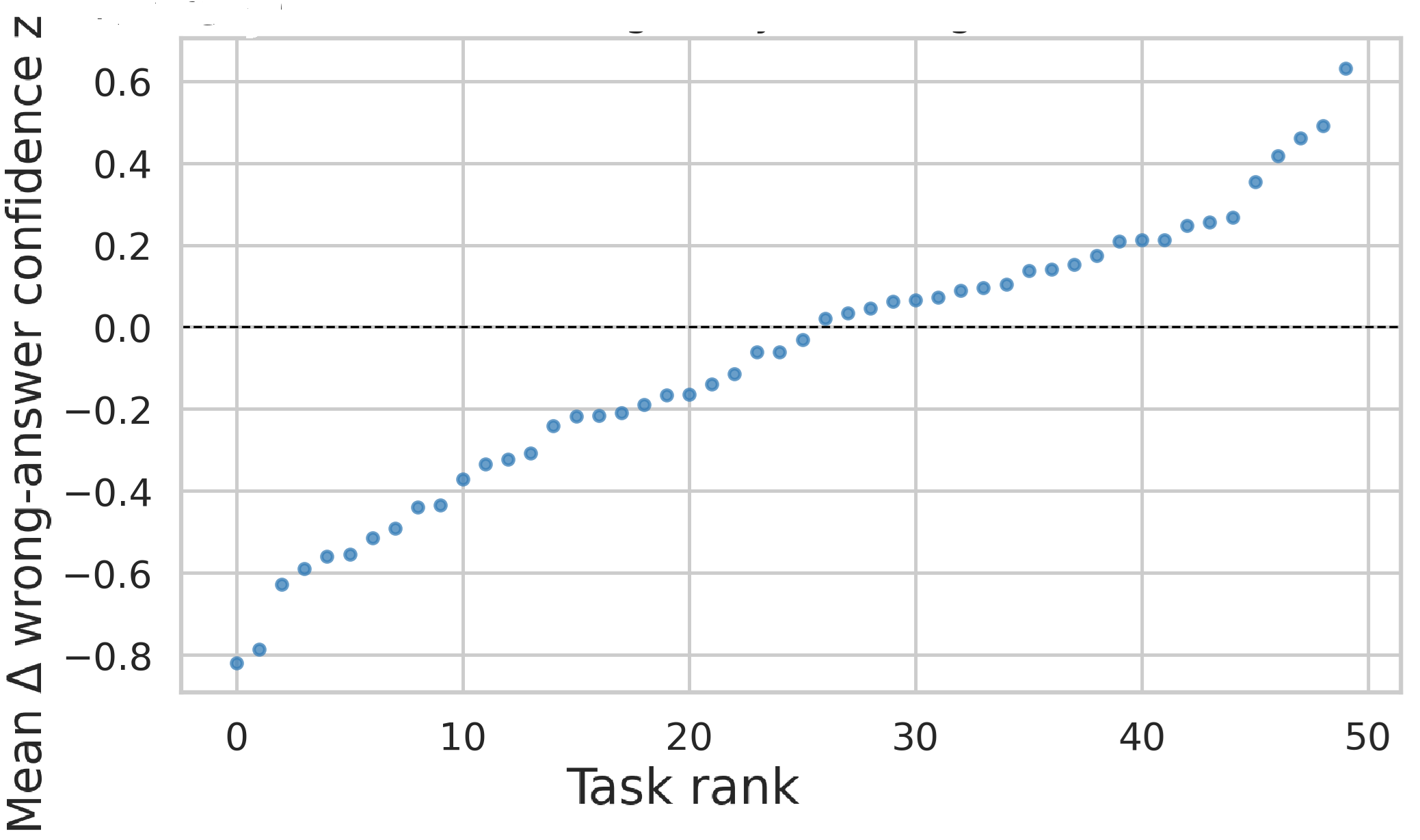}
    \vspace{-3pt}
    \caption{Per-task changes in wrong-answer confidence ($\Delta z = z_{\text{L1b}} - z_{\text{L1}}$), sorted by magnitude. Each point is a task; the dashed line indicates zero change.}    
    \label{fig:task_rank_wrong_answer_confidence}
\end{figure}

\begin{table}[t]
\caption{Counts of tasks with decreases ($\Delta z < 0$), increases ($\Delta z > 0$), and no change in wrong-answer confidence ($\Delta z = z_{\text{L1b}} - z_{\text{L1}}$), grouped by model.}
\centering
\small
\setlength{\tabcolsep}{4pt}
{
\scriptsize
%\vspace{-3pt}
\begin{tabular}{l l l c c c}
\hline
\textbf{Group} & \textbf{Type} & \textbf{Size} & \textbf{$\Delta z < 0$} & \textbf{$\Delta z > 0$} & \textbf{$\Delta z = 0$} \\
\hline
Overall (task mean) & All & All & \high{26/50} & \low{24/50} & 0/50 \\
smollm3-3b & Reasoning & Small & \high{25/36} & \low{11/36} & 0/36 \\
deepseek-r1-distill-qwen-1.5b & Reasoning & Small & \low{19/49} & \high{30/49} & 0/49 \\
deepseek-r1-distill-qwen-7b & Reasoning & Med. & \low{8/22} & \high{14/22} & 0/22 \\
phi-3.5-mini-instruct & Instruct & Small & \high{20/39} & \low{19/39} & 0/39 \\
qwen-0.5b & Instruct & Small & \high{27/44} & \low{17/44} & 0/44 \\
qwen-7b & Instruct & Med. & \high{18/30} & \low{12/30} & 0/30 \\
llama-3.1-8b-instruct & Instruct & Med. & \low{16/33} & \high{17/33} & 0/33 \\
qwen-7b-coder & Coder & Med. & \high{21/28} & \low{7/28} & 0/28 \\
deepseek-coder-6.7b & Coder & Med. & \high{30/49} & \low{19/49} & 0/49 \\
\hline
\end{tabular}
}
\label{tab:task_delta_z_counts}
\end{table}

Table~\ref{tab:figure3_localization_transition} reports combined effects on accuracy and HCI. Spike intensity affects accuracy in a displacement-dependent way — decreasing it under medium and high displacement, increasing it under low — with no consistent relationship. HCI, by contrast, rises with spike intensity for low and medium displacement and barely changes for high, pointing to identifier-level disruption as the more consistent driver of confident failures. Having examined both factors at moderate ranges, we next isolate tail subsets to test whether joint extremes exceed what either factor alone predicts.

Table~\ref{tab:spike_distance_subsets} compares performance within tail subsets, defined by high semantic displacement, high identifier-level disruption, and their intersection, against a low-distance/low-spike baseline. Semantic displacement alone has limited effect: the top-20\% distance subset shows accuracy comparable to baseline (33.11\% vs.\ 31.68\%) with a moderate HCI increase (6.89\% vs.\ 4.95\%). Identifier-level spikes alone produce a different pattern: accuracy rises to 36.00\% while HCI also increases to 8.44\%, pointing toward more decisive but less reliable predictions. These results localize semantic mis-framing to the \textit{atom level}. Identifier-level disruption (spikes) increases high-confidence errors, indicating that failures originate from instability around renamed identifiers rather than from structural reasoning.

When both factors co-occur, however, the interaction is non-additive and severe. Accuracy collapses to 21.25\%, a drop of 15 percentage points relative to the spike-alone subset, while HCI reaches 10.00\%, the highest value observed across all conditions. The sharpness of the accuracy collapse, rather than the modest HCI increment, is the clearest signal that joint extremes constitute a qualitatively distinct failure regime rather than an additive combination of two moderate effects. Whether these concentrated failures reflect consistent task-level patterns or are distributed unevenly across the evaluation set remains open; we examine this next.

We examine task-level variation by aggregating wrong-answer confidence changes across problems. Fig.~\ref{fig:task_rank_wrong_answer_confidence} shows the distribution of $\Delta z$ (L1b--L1) over persistently incorrect cases, ranging from $-0.8$ to $+0.65$, with no central concentration. This indicates that adversarial renaming does not produce a consistent directional effect on wrong-answer confidence. Table~\ref{tab:task_delta_z_counts} quantifies this heterogeneity: at the aggregate level the distribution is nearly balanced (26/50 tasks with $\Delta z < 0$ vs.\ 24/50 with $\Delta z > 0$), but model-level breakdowns reveal that this near-even split reflects opposing tendencies rather than a weak effect. Code-specialized models trend toward confidence suppression, while reasoning-distilled models show the opposite, indicating that the direction of the effect is jointly determined by task characteristics and model behavior.

\begin{tcolorbox}[colback=white, colframe=black, arc=8pt, boxrule=0.5pt]
    \textbf{RQ4 Takeaway:} \textit{Adversarial renaming triggers high-confidence wrong answers only when semantic displacement and identifier-level disruption occur together. The direction of confidence shifts is task- and model-dependent, marking this as a localized, atom-level failure rather than global structural degradation.}
\end{tcolorbox}

\section{Threats to Validity}

\emph{Contamination Control}. We use model release timing relative to our LeetCode dataset’s creation (May 2025) as a proxy to mitigate training data contamination. However, release dates are an imperfect indicator of training data cutoff, and some models post-date this threshold; thus, overlap with training corpora cannot be ruled out.
Datasets may limit generalizability.

\emph{Task and Evaluation Design}. We restrict the study to deterministic, self-contained snippets, which may bias results toward simpler programs and omit real-world complexities like external dependencies. Variants such as adversarial renaming may not fully represent obfuscations. 

{\em Evaluation Proxy}. Output prediction imperfectly proxies code understanding. We address this through various metrics.

\section{Related Work}

\emph{Obfuscation and Human Code Comprehension}. \cite{collberg1997taxonomy} provides a taxonomy motivating our tiered design. Surveys establish obfuscation as a controlled probe of model robustness~\cite{schrittwieser2016protecting,banescu2017predicitngresilience}.
Identifier renaming significantly reduces task performance~\cite{ceccato2014family}, and program comprehension depends on working memory, attention, and semantic cues~\cite{siegmund2014understanding,siegmund2017measuring}. Meaningful identifiers reduce cognitive load; poor naming increases it~\cite{lawrie2006identifiers,binkley2013impact}. Most directly, Nguyen \textit{et al.}~\cite{nguyen2026effectcodeobfuscationhuman} measured human comprehension under these obfuscation tiers; we build on their paradigm and data to study whether LLMs exhibit the same difficulty patterns.

\emph{LLM Reasoning and Cognitive Alignment}. 
CoT prompting improves performance on difficult tasks \cite{wei2022chainofthought,wang2023selfconsistency,kojima2022largelanguagemodelszeroshot}, and reasoning effort scales with task difficulty \cite{devarda2025cost,yan2024chatgptdifficulty}. Reasoning traces may be post-hoc rather than faithful \cite{turpin2023languagemodelsdontsay,lanham2023measuringfaithfulnesschainofthoughtreasoning}. Token-level uncertainty is linked to confusion in humans/models \cite{abdelsalam2025humansllmsprocessconfusing}. 

Researchers capture human attention from eye movements and use it to improve neural code summarization~\cite{bansal2023humanattention}. EyeTrans~\cite{zhang2024eyetrans} integrates human attention into Transformer-based code summarization. More recent work further explores training code models to mimic human visual attention~\cite{zhang2025eyemulator}. The attention weights of neural models of code are compared with human visual attention during code summarization, asking whether models attend to the same tokens as developers~\cite{paltenghi2021thinking}.

\emph{Expert--Novice Differences in Program Comprehension}. Experts rely on plan-based and hierarchical reasoning, whereas novices tend to process code locally and line by line~\cite{soloway1984empirical,fix1993expert}. Pennington~\cite{pennington1987stimulus} shows that expert understanding integrates control-flow and goal-oriented representations, while Corritore and Wiedenbeck~\cite{corritore2001exploratory} identify identifiers as semantic ``beacons.'' 

\emph{Identifier Sensitivity and Adversarial Robustness in Code Models}.
Identifiers carry rich semantics in programs~\cite{allamanis2015suggesting,allamanis2018learningrepresentprogramsgraphs}, yet adversarial renaming misleads models~\cite{zhang2020generating,yefet2020adversarialexamples,henkel2022semantic}. Code models often rely on surface cues over deep semantics~\cite{rabin2021onthegeneralizability}, and identifier quality affects performance~\cite{nikiema2025codebarrierllmsactually}.

\section{Implications and Conclusion}
\label{sec:implications}

\subsection{\bf \em Implications for software engineering}
\label{subsec:se}

1. Our findings bear on {\bf \em human-AI collaboration} in SE. Tasks like code review, debugging, and comprehension require developers to interpret and act on model outputs, so model selection should weigh not only accuracy but whether {\em the model's reasoning aligns with how developers understand code}. Reasoning-tuned models are both more accurate and {\em more aligned with developers'} difficulty patterns, making {\bf \em human alignment an additional criterion} for developer-facing tools: a model whose difficulty profile resembles a developer's yields reasoning traces and failure modes easier to inspect, calibrate, and supervise, especially when explaining unfamiliar code, assisting
debugging, or code comprehension and review.

2. Our results also expose a concrete risk: {\bf \em adversarial or suspicious naming can cause high-confidence incorrect answers} when semantic displacement and identifier-level disruption co-occur. Since developers often face poorly named, generated, minified, legacy, or unfamiliar code, tools should not treat confident explanations as evidence of understanding when identifier semantics are suspicious; such cases should trigger extra checks like running, testing, or review.

3. Our results also caution against treating verbose reasoning as reliable: very long CoT traces show diminishing returns, so {\bf \em length past a moderate budget is a weak reliability~signal} under stronger obfuscation. Unusually long traces, repeated self-correction, or uncertainty around identifiers should instead be read as signs of a struggling model, warranting extra checks.

\subsection{\bf \em Implications for software security}

1. Obfuscation is widely used to protect software from reverse engineering, code pirating, and tampering.
%and is common in malware, packed scripts, and minified code. 
If LLMs struggle with obfuscated code as humans do, obfuscation stays relevant even under LLM-assisted analysis: our findings suggest {\bf \em current LLMs do not eliminate its protective value}, as {\em  transformations like adversarial renaming or control-flow flattening still create meaningful comprehension~barriers}.

2. Our findings also identify a concrete risk for {\bf \em malware and malicious code}: {\bf \em adversaries may intentionally choose benign-looking identifiers to mislead LLM-based analyzers}, leading a model to confidently infer safe behavior while missing malicious functionality. As in the SE setting (\S\ref{subsec:se}), confident explanations are not sufficient evidence of safety when identifier semantics are suspicious. The shared failure patterns between humans and LLMs also identify {\bf \em security-relevant blind spots}. Adversarial
renaming can confuse both humans and models, causing models to produce high-confidence incorrect answers.

3. Not all models fail equally: reasoning-tuned models are more robust and human-aligned, suggesting {\bf \em they may serve as analyst-facing assistants for malware analysis and vulnerability triage}. Their value is not to replace analysts but to fail more predictably, making them easier to supervise---and in security workflows, where a wrong explanation can mean missed malicious behavior or incorrect assessment, human-interpretable failure is as crucial as accuracy.\\
\noindent {\bf \em \underline{Data Availability Statement}}: Our package is available at \cite{anon_data_repo_2026}.

%\vspace{2pt}
%\noindent {\bf \em \underline{Data Availability Statement}}: Our package is available at \cite{anon_data_repo_2026}.

\newpage

\bibliographystyle{IEEEtran}

\bibliography{references}

\end{document}